\documentclass{aa}
\usepackage{graphicx} 
\usepackage{txfonts}
\usepackage{amsmath}
\usepackage{amsfonts}
\usepackage{hyperref}
\usepackage{bm}
\usepackage{subcaption}
\usepackage{xcolor}
\usepackage{soul} 
\usepackage{natbib}
\bibpunct{(}{)}{;}{a}{}{,}

\hypersetup{
colorlinks = true, 
linkcolor = blue, 
citecolor = blue, 
filecolor = magenta, 
urlcolor = blue
}

\def \mancha{{\textsc{Mancha3D~}}}

\begin{document}

   \title{Solar vortex detection methods in MHD simulations: impact of magnetic field and spatial resolution}

   \author{M.~Koll~Pistarini\inst{1, 2} \and  E.~Khomenko\inst{1, 2} \and T.~Felipe\inst{1, 2} \and M.~Modestov\inst{1, 2, 3}}

   \institute{Instituto de Astrofísica de Canarias, 38205 La Laguna, Tenerife, Spain          \and
    Departamento de Astrofísica, Universidad de La Laguna, 38205 La Laguna, Tenerife, Spain  \and
    Institute of Chemical Kinetics and Combustion SB RAS, 630090 Novosibirsk, Russia
             }

  \abstract
   {Small-scale vortices are ubiquitous structures that exist over a wide range of spatial extents in the solar atmosphere. They are considered to play an important role in the energy transport and local heating. However, numerical simulations frequently assess the effect of vortices without simultaneously considering the effects of spatial resolution and differences in magnetic configuration.}
   {We investigate the influence of different magnetic field configurations and spatial resolutions on vortex structures. We evaluate two detection methods to determine their impact on the morphology, statistical properties, and temperature profiles of vortices.}
   {We analyzed a set of six three-dimensional realistic simulations of the solar atmosphere under three different magnetic field configurations: a small-scale dynamo and two initially vertical implanted magnetic fields of $50$ G and $200$ G. Three different spatial resolutions have been employed: $20 \times 20 \times 14$, $10 \times 10 \times 7$ and $5 \times 5 \times 3.5$ km$^3$. We applied two vortex detection methods based on the velocity gradient tensor to all of the models: swirling strength and the SWIRL code. We performed a comparison of vortex locations obtained with both methods, and a statistical analyses of the vortex generation mechanisms, the area covered by vortices, their number and characteristic sizes, and temperature profiles as a function of height.}
   {We have confirmed that different magnetic field configurations and spatial resolutions impact the area coverage, number, and sizes of vortices. Likewise, the detection methods impact the statistics obtained. Swirling strength detects vortices with any orientation but a height-dependent threshold is needed. SWIRL only detects vertically-oriented vortices but shows a better agreement with the rotating horizontal velocity field. Simulations with a vertical magnetic field of $50$ G support the formation of chromospheric vortices without a photospheric counterpart, while most of the vortices in the $200$ G model directly connect the photosphere with the chromosphere. Small-scale dynamo simulations are characterized by a large number of horizontal vortices, with vertical vortices being nearly absent at chromospheric layers. Temperature profiles of vortices confirm that they are hotter than their surroundings, regardless the simulation setup.} 
   {Vortices are ubiquitous throughout the solar atmosphere. Their detection is affected by the methodology employed, with SWIRL being a more robust method. The strength of the magnetic field and the numerical resolution considerably affect the number and size of the structures detected. Despite this, they all show temperatures higher than those of their surroundings. }

            
   \maketitle

   \nolinenumbers

\section{Introduction}

Over the last decade, vortical structures have been proposed as important mechanisms for transporting energy throughout the solar atmosphere. In the photosphere, vortical motions have been identified in intergranular lanes by tracking the motion of magnetic bright points \citep{brandt1988, bonet2008} or by inferring the horizontal velocity field with Local Correlation Tracking (LCT) methods \citep{bonet2010, vargasdominguez2011}. Subsequently, several detections have been reported at chromospheric layers, where swirling motions are primarily identified by following morphological signatures across high-resolution time series \citep{wedemeyer-bohm2009, park2016, tziotziou2018, dakanalis2022}. \citet{wedemeyer-bohm2012} performed a multi-wavelength study combing data from SST/CRISP and SDO/AIA. For the first time, the authors found signatures of a vortical structure that connects the photosphere with the solar corona, suggesting that they could act as channeling conduits that transport energy along them. 

Current vortex observations in quiet Sun regions show a wide range of spatial scales. Large-scale structures with diameters of approximately $2-5$ Mm have been identified in the photosphere \citep{brandt1988, balmaceda2010, requerey2018} and the chromosphere \citep{wedemeyer-bohm2009, morton2013, dakanalis2022}. Reported small-scale vortices are often limited by the spatial resolution of the current instrumentation. However, features of only about $0.5$ Mm in diameters have been observed in both layers \citep[e.g.,][]{vargasdominguez2011, requerey2017, giagkiozis2018, park2016, liu2019}.

Realistic MHD numerical simulations allow us to study vortical structures at a level of detail and spatial resolution that is not possible to achieve with current observations. In addition, the complete information of the velocity and magnetic vector fields is available at any time instants. Several works have focused on the analysis of vortices at photospheric and chromospheric heights \citep[e.g.,][]{moll2011, moll2012, kitiashvili2011, yadav2020, yadav2021, silva2020, silva2021, khomenko2021}. The analysis of vortices have been performed using different numerical codes such as \textsc{Bifrost~} \citep{gudiksen2011}, \textsc{CO$^5$BOLD~} \citep{freytag2012}, \mancha \citep{modestov2024} or MURaM \citep{vogler2005, rempel2014, rempel2017}. However, small differences in the internal implementation of the numerical schemes can affect the comparison of simulations computed with different codes \citep{fleck2021}.

Most vortex studies analyze a single magnetic field configuration, with simulations that include vertically implanted magnetic fields between $10$ to $200$ G, as well as dynamo models \citep[for a detailed summary see][]{tziotziou2023}. \cite{battaglia2021} and \cite{kannan2024} included a comparison between three magnetic field configurations, finding differences in vortex generation among magnetizations. They suggest that intermediate magnetic fields of about $50$ G can favor the generation of chromospheric vortices due to a balance between magnetic and gas pressure. 

Previous works have analyzed the equation of vorticity \citep{stein1998, shelyag2011, silva2024b} to study the main drivers of vortex generation in a single magnetic field configuration. But as \citet{canivetecuissa2020} showed, this equation is not the best choice as it includes shear flows. They proposed a new equation of evolution of the swirling strength which is not affected by them. A detailed analysis of the main vortex generation mechanisms using this new approach is needed to assess the impact of different magnetic field configurations without being contaminated by shear flows. 

In addition, a variety of spatial resolutions have been used, typically ranging from $10$ to $20$ km in all spatial dimensions. While the study by \citet{moll2011} included three models at different spatial resolutions, the authors did not analyze the impact of spatial resolution on the identified vortex. To our knowledge, no previous studies have directly analyzed this effect.

Typical small-scale vortex sizes in magneto-convection simulations have diameters of around $50-100$ km in the photosphere, reaching values of $100-200$ km in the chromosphere \citep{silva2020, yadav2021, cuissa2023}. There are few detections in numerical simulations of larger vortices with diameters between $1-2$ Mm \citep{yadav2020, silva2024a}, but their rate of occurrence appears to be lower compared to the smallest vortices \citep{tziotziou2023}. Simulations of coronal loops have also revealed a few detections of larger vortices \citep{breu2023, kuniyoshi2025}. \cite{yadav2020} observed that most of the energy is transported by small-scale vortices, significantly contributing to the upward Poynting flux due to horizontal motions. Therefore, further investigation of small-scale vortices is required, given their abundance and energetic contribution to the solar atmosphere.

A key aspect in vortex analysis is the detection method employed. Despite it is easy to understand the appearance of vortices in a vector field, a mathematical definition of this phenomenon is rather complex. Identifying these structures in turbulent and compressible flows further increases the complexity, being one of the long-standing problems in fluid dynamics \citep{jeong1995, gunther2018}. In the last years, many efforts have been dedicated to the development of new techniques for vortex detection. In solar physics, several detection methods have been applied: vorticity \citep{stein1998, shelyag2011}, $\Gamma$-Functions methods \citep{graftieaux2001, giagkiozis2018}, Lagrangian-Averaged (LAVD) and Instantaneous (IVD) Vorticity Deviations \citep{haller2016, silva2020, silva2021} or velocity gradient tensor criteria such as Q-criterion \citep{khomenko2021, silva2021}, swirling strength \citep{moll2011, moll2012,yadav2020,yadav2021} or Rortex methods \citep{cuissa2022}. Each method allows to isolate vortical motions to some extent, although they have specific advantages and limitations depending on the research objectives.

In this paper, we aim to perform a complete statistical analysis of vortical structures in numerical simulations. In particular, we investigate how vortical structures are affected across different magnetic field configurations and spatial resolutions by analyzing source terms of generation of vorticity, morphological properties and temperature profiles. For this purpose, we analyzed \mancha simulations comparing two vortex detection methods: the swirling strength criterion \citep{zhou1999} and the recently developed SWIRL code \citep{cuissa2022}. 

The paper is organized as follows: Section~\ref{sec:Numerical_simulations} describes the numerical setup, while Section~\ref{sec:detection_methods} provides a description of the two detection methods used. Our main results are shown and discussed in Section~\ref{sec:results_discussion}. Conclusions are presented in Section~\ref{sec:conclusions}.

\section{Numerical datasets}
\label{sec:Numerical_simulations}

\renewcommand{\arraystretch}{1.4} 
\begin{table}
\caption{Total duration of the simulation runs analyzed.}
\centering
\begin{tabular*}{0.9\columnwidth}{@{\extracolsep{\fill}}cccc}
\hline\hline
$\Delta x \times \Delta y \times \Delta z \; \text{km}^3$ & SSD & Bz50 & Bz200 \\
\hline
$20 \times 20 \times 14$ & $20$ min & $26.3$ min & $17.3$ min \\
$10 \times 10 \times 7$ & $20$ min & $14.5$ min & - \\
$5 \times 5 \times 3.5$ & $4$ min & - & - \\
\hline
\end{tabular*}
\label{tab:sims_summary}
\end{table}
\renewcommand{\arraystretch}{1}

We used of a set of realistic three-dimensional radiative MHD simulations computed with the \mancha code \citep{modestov2024}. The code solves the single-fluid, non-linear MHD equations and can account for non-ideal processes such as ambipolar diffusion, Hall term, and Biermann battery effect. The set of MHD equations is closed with a realistic equation of state based on solar abundances \citep{anders1989}. Radiative losses are computed under Local Thermodynamic Equilibrium (LTE) using the gray approximation. The computational domain of all simulations spans $5.76 \times 5.76 \times 2.3$~Mm$^3$, covering from about $0.9$~Mm below the photosphere up to $1.4$~Mm into the low chromosphere. Periodic lateral boundaries are applied. The bottom boundary is open with controlled conditions for mass and energy fluxes to reproduce the solar radiative flux. The upper boundary is closed for mass flows and symmetric conditions (zero gradient) are applied for density and internal energy. Further details about the calculation of the simulations can be found in \cite{khomenko2025}.

Three different magnetic configurations are available: a small-scale dynamo (SSD), where the magnetic field is seeded by the Biermann battery term \citep{khomenko2018}; and two initially implanted uniform vertical magnetic field of $50$~G (Bz50) and $200$~G (Bz200), which are representative of a medium magnetized quiet-Sun region and a plage area, respectively. Simulations with three different spatial resolutions have been computed: $20 \times 20 \times 14$ km$^3$ (low), $10 \times 10 \times 7$ km$^3$ (medium), and $5 \times 5 \times 3.5$ km$^3$ (high), covering the same computational domain. SSD simulations are available at all three spatial resolutions, whereas the Bz50 model was computed at medium and low resolutions, and the Bz200 simulation only at the lower resolution. The temporal evolution of all simulations is about $20$ minutes of solar time, with snapshots saved every $10$ seconds. The SSD model at high spatial resolution is only available for $4$ minutes due to computational constraints. Table \ref{tab:sims_summary} shows a summary of the six simulation setups. The models analyzed in this study do not include non-ideal effects, with the exception of the Biermann battery term in the SSD models (they correspond to the \textit{clean} models described in \cite{khomenko2025}).

\section{Vortex detection methods}
\label{sec:detection_methods}

In this study, we applied two vortex detection methods to evaluate their performance on our set of numerical simulations: the swirling strength criterion \citep{zhou1999} and the SWIRL code \citep{cuissa2022}. Swirling strength has been widely used in the literature to detect and analyze solar vortices in numerical simulations \citep[e.g.][]{moll2011,moll2012,kato2017,yadav2020,yadav2021,battaglia2021,kannan2024}. In contrast, SWIRL is a novel method with an advanced detection process, which has only been used by \citet{cuissa2023}. Applying these methods allowed us to systematically compare their results across different models, as both are based on the velocity gradient tensor.

\subsection{Swirling strength} \label{sec:swirling_strength}

This method is based on the decomposition of the velocity gradient tensor $\mathcal{U}_{ij} := \partial_{j} v_{i}$ into eigenvectors and eigenvalues. Vortex structures are located in regions where the velocity gradient tensor has a complex conjugated pair of eigenvalues \citep{chong1990}. Then, $\mathcal{U}$ can be decomposed as follows,
\begin{equation}
\mathcal{U} = 
\underbrace{
  [\bm{u}_{\text{r}}, \bm{u}_{+}, \bm{u}_{-}]
  \vphantom{\begin{bmatrix} 0\\0\\0 \end{bmatrix}}
}_{\text{\large $\mathcal{P}$}} 
\underbrace{
  \begin{bmatrix}
  \lambda_{\text{r}} & 0 & 0 \\
  0 & \lambda_{c+} & 0 \\
  0 & 0 & \lambda_{c-}
  \end{bmatrix}
}_{\text{\large $\Lambda$}} 
\underbrace{
  [\bm{u}_{\text{r}}, \bm{u}_{+}, \bm{u}_{-}]^{-1}
  \vphantom{\begin{bmatrix} 0\\0\\0 \end{bmatrix}}
  }_{\text{\large $\mathcal{P}^{-1}$}} \, ,
\end{equation}
where $\lambda_r$, $\lambda_{c+}$  and $\lambda_{c-}$ are the eigenvalues associated with the eigenvectors $\mathbf{u}_r$, $\mathbf{u}_+$, $\mathbf{u}_-$, respectively. The direction of the rotation axis is given by the real eigenvector $\mathbf{u}_r$, while $\mathbf{u}_+$ and $\mathbf{u}_-$ are the eigenvectors that characterize the swirling motion in the rotation plane \citep{zhou1999}. $\mathcal{P}$ refers to the matrix composed by the eigenvectors, with $\mathcal{P}^{-1}$ its inverse, and $\Lambda$ represents the matrix containing the eigenvalues. The complex conjugated pair of eigenvalues ($\lambda_{\mathrm{c}\pm}$) can be expressed as $\lambda_{\mathrm{c}\pm}  = \lambda_{\mathrm{cr}} \pm i\lambda_{\mathrm{ci}}$, where $\lambda_{\mathrm{cr}}$ and $\lambda_{\mathrm{ci}}$ are the real and imaginary parts, respectively. The strength of the vortex rotation (\textit{swirling strength}) can be characterized by the unsigned imaginary part $\lambda_{\mathrm{ci}}$ \citep{zhou1999}. In addition, it is possible to define a swirling strength period ($\tau_{ci}$), which characterizes the period of rotation of swirling structures by the relation $\tau_{ci} = 2\pi/\lambda_{ci}$.

Swirling strength enables the detection of rotational motions from the three-dimensional velocity field, which allows to trace vortex regions oriented in any spatial direction. However, it requires to select a threshold to remove weak vortices or noisy regions not associated with vortices. In a turbulent plasma, the choice of the threshold should be done by visual inspection to ensure reliable results. A comparison between the spatial distribution of swirling strength with the horizontal velocity field allows to establish the optimal threshold that better isolates vortex structures.

A common approach consists in applying a fixed threshold to the whole computational domain \citep[see e.g.][] {moll2012, battaglia2021, yadav2021}. However, since the threshold depends on the amplitude of the flow, it might not remain accurate in simulations spanning a wide range of heights in a gravitationally stratified atmosphere. Another approach is to apply a height-dependent threshold based on the velocity field at each height. In this way, it is not necessary to rely on a visual comparison to determine an adequate threshold. We applied the following definition: $\lambda_{\text{th}} =  \mu \, + \, 0.5\sigma$, where $\mu$ and $\sigma$ are the mean value and standard deviation of the logarithmic $\lambda_{ci}$ distribution at each height, respectively \citep[similarly to][]{kannan2024, kuniyoshi2025}. We only consider vortex regions those that satisfy $\lambda_{ci} > \lambda_{\text{th}}$ at each height.

Grid points associated with vortex structures are independently computed at each time instant. The detections are expected to present some degree of temporal coherence and, ideally, this information should be included in the analysis. However, addressing this issue in small-scale vortices is a challenging task. The method does not provide isolated individual vortices, which hinders our ability to track the same vortex structure in time. Thus, in this method we do not take into account the temporal coherence of the detections to filter out transient structures.

\subsection{SWIRL code} \label{sec:SWIRL}

The SWIRL code is a newly developed open-access tool for automatic detection of vortices \citep{cuissa2022}. It is based on the mathematical quantity Rortex \citep{tian2018, liu2018}, which is also related to the velocity gradient tensor. Rortex is derived by expressing the velocity gradient tensor in a reference frame aligned with the rotational axis of the fluid given by $\mathbf{u}_r$ (see Sec.~\ref{sec:swirling_strength}). The magnitude of Rortex is directly linked to the pure rotational component, making it a more robust method compared to swirling strength, as it fully decouples contributions from shear motions. However, a key limitation of Rortex lies in its computational complexity because it requires the calculation and decomposition of several matrices. To address this issue, \citet{wang2019} and \citet{xu2019} derived a new expression for Rortex computation, significantly reducing the associated computational costs:
\begin{equation} \label{eq:Rortex}
R = \bm{\omega} \cdot \bm{u_{r}} - \sqrt{ \left( \bm{\omega} \cdot \bm{u_{r}} \right)^{2} - (2\lambda_{ci})^{2} } \; , 
\end{equation}
where $\bm{\omega}=\nabla \times \bm{v} $ is the vorticity. SWIRL combines the Rortex quantity with a clustering algorithm \citep{rodriguez2014} to automatically identify vortex locations. It estimates and groups local centers of curvature from the velocity field. Regions with high density concentration of points are indicative of potential vortex locations. The clustering approach enhances the detection of regions with a closed velocity field.

The SWIRL code takes a $2$D velocity field as input. We provided the horizontal components of the velocity field ($v_x$, $v_y$). This approach allows for a better comparison with observations, which are mostly restricted to the horizontal plane. However, the code can be applied to any other pair of components of the velocity field. Additionally, SWIRL requires a parameter file which must be tuned to the specific simulation setup (see Appendix \ref{app:SWIRL_params}). We applied the same parameter file across all heights for a given model. Simulations with different magnetic field configurations and spatial resolutions require small adjustments of the parameters to optimize the performance of the code. The most important parameter to adjust is the set of stencils used for the computations (see Appendix \ref{app:SWIRL_params} for details). These stencils control the scales at which curvatures in the velocity field are detected through the calculation of the velocity gradient tensor derivatives. This represents a significant advantage over the swirling strength criterion, which relies on a single scale.

The output consists of grid cell locations where vortices are detected, with the information for each vortex being individually separated. We removed vortices with less than $5$ grid points as we consider that there is not enough information to assure that they are rotating structures. We refined the area associated to each vortex by imposing that all grid points belonging to a single structure form a closed contour. To that end, we applied a convex hull algorithm \citep{barber1996} to each vortex detection. This improves the quality of the results by providing more circular shaped contours that better fit the rotating velocity field. The process only adds a few grid points to the area of each vortex, without affecting the locations and centers of the detections. Since we have each structure isolated, temporal information can be included across consecutive snapshots, unlike the swirling strength method (Sec.~\ref{sec:swirling_strength}). We applied the condition that detections must persist at a close distance for at least $10$ seconds before and after the snapshot under consideration. This is done by checking if at least one grid point of the vortex is shared by the vortex detections in the previous and posterior snapshots. Thus, we ensure that the analyzed structures have a minimum lifetime of $20$ seconds (meaning three consecutive snapshots). The applied selection helps to filter out possible false detections and transient structures identified by the code. To perform the three-dimensional analysis of vortices, the code was applied to all heights of the simulation domain, allowing the reconstruction of the complete vortex structures. We consider a vortex to belong to the same three-dimensional structure as long as the separation between the detected centers in the vertical and horizontal directions is less than $3$ grid cells in each direction for the low resolution models. This value is scaled by the resolution improvement factor to maintain consistent physical structures across different resolutions.

The results include a wide variety of rotating regions with different sizes and rotation speeds. Thus, the code can identify vortices with weak rotation speeds in comparison with the average velocities at the same height. We are interested in vortices rotating fast enough to meaningfully contribute to the energy transport. Thus, vortices with very low rotation speeds compared to the average velocity flows have been removed. Indeed, this selection is analogous to the application of a threshold in the swirling strength method (see Sec. \ref{sec:swirling_strength}), which excludes weaker vortex motions. To address this issue, we used the Rortex quantity computed by the SWIRL code with the smallest stencil (see Appendix \ref{app:SWIRL_params}). We restrict the study to vortices that present Rortex values above the mean value at the corresponding height in at least $10\%$ of the detected vortex area. This criterion has been determined by visual inspection through a trial and error process in all of the simulations (see Appendix \ref{app:SWIRL_Rortex_filter} for a visual example).

\section{Results and discussion}
\label{sec:results_discussion}

\subsection{Swirling strength threshold} \label{sec:ss_threshold}

\begin{figure}[t]
    \centering
    \begin{subfigure}[h]{0.49\columnwidth}
        \includegraphics[width=\columnwidth]{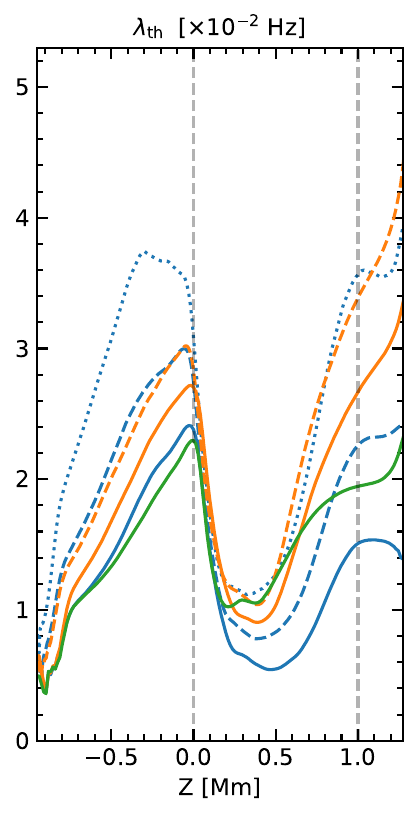}
    \end{subfigure}
    \begin{subfigure}[h]{0.49\columnwidth}
        \includegraphics[width=\columnwidth]{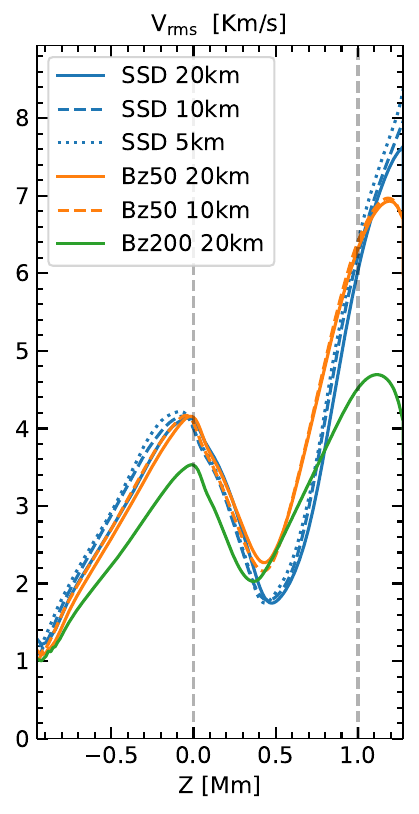}
    \end{subfigure}
    \caption{Horizontal and temporal average of the height-dependent swirling strength threshold $\lambda_{\text{th}}$ (left) and the rms velocity (right). The meaning of the curves is indicated in the figure. Vertical gray dashed lines are references at $Z=0$ Mm and $Z=1$ Mm.}
    \label{fig:v_rms_and_ss_var_criterion}
\end{figure}

Left panel of Fig.~\ref{fig:v_rms_and_ss_var_criterion} shows the spatio-temporal averages of the height-dependent threshold as a function of height (see Sec.~\ref{sec:swirling_strength}). This quantity is related to the average rotation speeds in vortex regions. We observe a similar behavior between models but with differences in magnitude. In the convection zone (below $Z=0$ Mm), results are essentially grouped by spatial resolution. Higher resolution simulations improve the accuracy of the velocity gradient tensor and enable the detection of small vortex structures that would remain unresolved at lower resolutions. This leads to an increase in swirling strength magnitude. At the solar surface, $\lambda_{\text{th}}$ quantity decreases and converges for all the models. Differences are amplified again in the chromosphere. At the same spatial resolution, Bz50 simulations yield vortices that exhibit faster rotation speeds, followed by Bz200 and SSD models.

We compute the rms velocity profiles of each model (right panel of Fig.~\ref{fig:v_rms_and_ss_var_criterion}) to evaluate its relationship with the height-dependent threshold $\lambda_{\text{th}}$. The velocity profiles show the dynamics of the simulations across the domain. In the convection zone, the velocities increase with height up to the solar surface, where convection stops and plasma motions are slowed down. There, low amplitude waves are excited, which slowly increase their amplitude with height due to the drop in density. Thus, velocity perturbations grow up to the top of the domain. This behavior in the velocity profiles has been previously reported in the literature \citep{fleck2021}. All models show a similar behavior. Bz200 simulation exhibits a lower velocity profile compared with the other models, as strong magnetic fields suppress plasma motions. We do not find significant differences between spatial resolutions.

Comparing both panels of Fig.~\ref{fig:v_rms_and_ss_var_criterion} reveals that the height-dependent threshold is intrinsically linked to the rms velocity profiles. This is expected as swirling strength is derived from the velocity field. Both quantities exhibit a strong dependence on atmospheric height, highlighting the importance of applying a height-dependent threshold. Each model requires a specific selection of values. Bz50 simulations allow for vortices with faster average rotation speed, despite having velocity profiles comparable to the SSD models. Stronger magnetic fields suppress plasma motions, leading to a decrease in vortex rotation speeds. Vortices in weaker magnetic fields configurations are found in slower plasma flows compared with unipolar models. Therefore, these results suggest that intermediate magnetic fields are optimal for generating vortices with enhanced rotation (as also observed by \cite{battaglia2021} and \cite{kannan2024}). Increasing the spatial resolution also leads to higher swirling strength values, mainly in the convection zone and the chromosphere.

\subsection{Swirling strength generation terms} \label{sec:ss_gen_terms}

\begin{figure}[t]
    \centering
    \includegraphics[width=0.9\columnwidth]{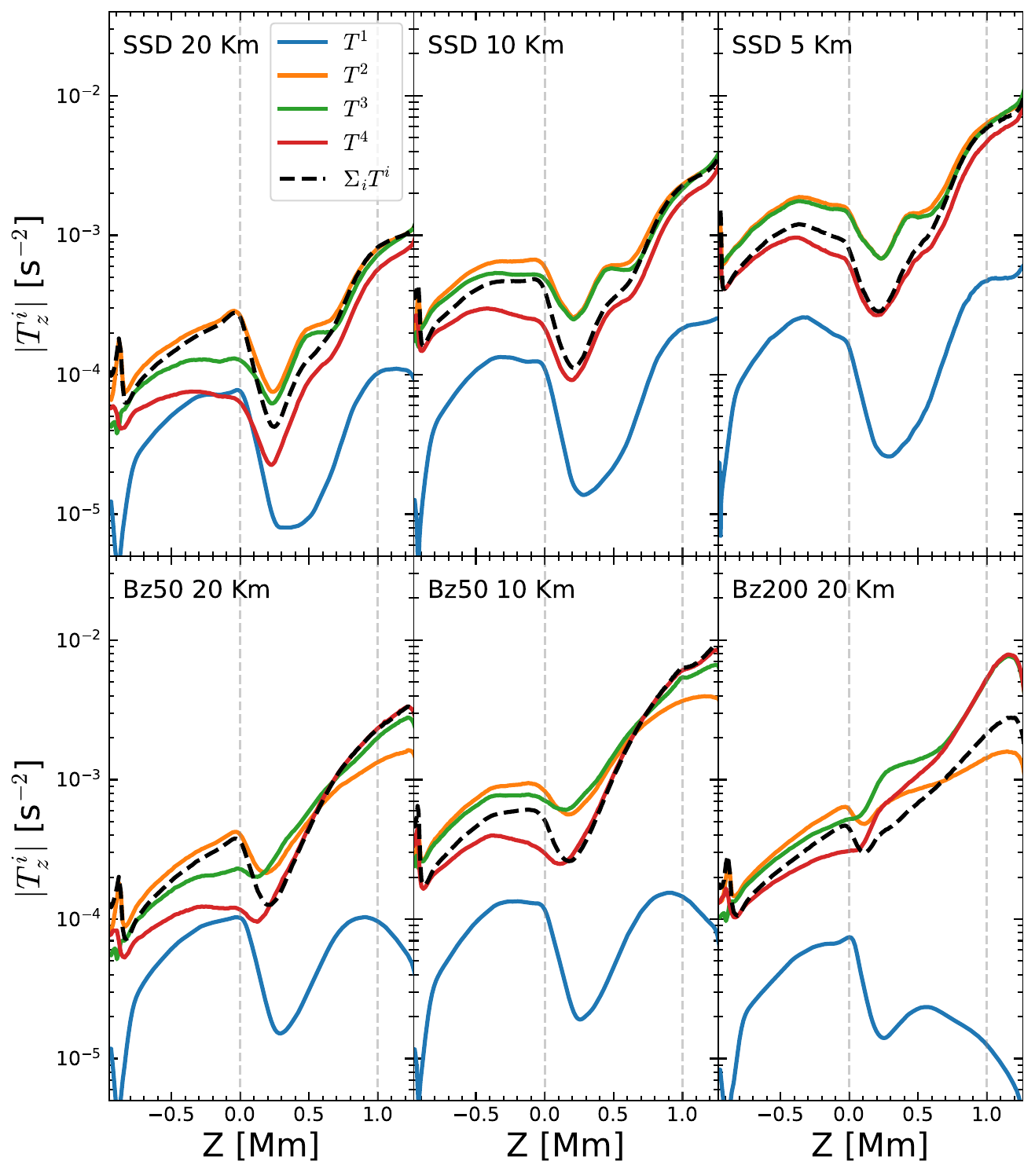}
    \caption{Horizontal and temporal averages of the absolute values of the vertical component of the swirling strength generation terms given by Eq.~\ref{eq:ss_terms} as a function of height, corresponding to the hydrodynamic stretching ($T^1$, blue lines) and baroclinic ($T^2$, orange lines) terms, and the magnetic baroclinic ($T^3$, green lines) and tension ($T^4$, red lines) terms. The black dashed line represents the total production of swirling strength given by $\sum_i T^i$. Each panel corresponds to the model indicated in the top left corner. Vertical gray dashed lines are visual references at Z $= 0$ Mm and Z $= 1$ Mm. }
    \label{fig:ss_generation_terms}
\end{figure}

Analyzing the equation of evolution of vorticity allows to understand the main mechanisms of vortex generation. Several previous studies apply the equation of evolution of vorticity $\omega$ \citep{stein1998, shelyag2011, silva2024b}. But this equation is not entirely suitable for studying the mechanisms that generates vorticity, as it also includes shear flows. Following the derivation of \citet{canivetecuissa2020}, it is possible to compute the time evolution of swirling strength as:
\begin{equation} \label{eq:ss_terms}
\begin{aligned}[b]
\frac{d}{dt}\lambda_{\mathrm{ci}} &= -2\lambda_{\mathrm{ci}}\lambda_{\text{cr}} && T^1 \\[5pt]
&\quad - \text{Im} \left\{ \mathcal{P}^{-1} \left[ \nabla \otimes \left( \frac{1}{\rho} \nabla p_{\text{g}} \right) \right] \mathcal{P} \right\}_{22} && T^2 \\[5pt]
&\quad - \text{Im} \left\{ \mathcal{P}^{-1} \left[ \nabla \otimes \left( \frac{1}{\rho} \nabla p_{\text{m}} \right) - \left( \nabla \frac{1}{\rho} \right) \otimes (\boldsymbol{B} \cdot \nabla)\boldsymbol{B} \right] \mathcal{P} \right\}_{22} && T^3 \\[5pt]
&\quad + \text{Im} \left\{ \mathcal{P}^{-1} \left[ \frac{1}{\rho} \nabla \otimes \left( (\boldsymbol{B} \cdot \nabla)\boldsymbol{B} \right) \right] \mathcal{P} \right\}_{22} && T^4 \\[5pt]
&\quad - \text{Im} \left\{ \mathcal{P}^{-1} \left[ \nabla \otimes (\nabla \Phi) \right] \mathcal{P} \right\}_{22}. && T^5
\end{aligned}
\end{equation}
where $\rho$, $p_{\text{g}}$, $p_{\text{m}}$, $\bm{B}$, and $\Phi$ are the plasma density, the gas and magnetic pressure, the magnetic field vector, and the potential of conservative forces (gravity in our case), respectively. The symbol $\otimes$ represents the tensor product, meaning that the result is a $3 \times 3$ matrix, from which we are only interested in the imaginary part of the $(2,2)$ component \citep[for further details see][]{canivetecuissa2020}. 

Each term of Eq.~\ref{eq:ss_terms} has a physical interpretation analogous to that of the terms in the vorticity equation. $T^1$ and $T^2$ are related to hydrodynamical processes, and they are interpreted as the stretching and baroclinic terms, respectively. $T^3$ and $T^4$ account for the magnetic contributions to the swirling strength generation. The former is related to the magnetic baroclinicity while the latter represents the magnetic tension. $T^5$ accounts for the generation of swirling strength due to the effect of conservative forces. As we employ a constant gravitational field across the domain, $T^5$ will not be considered further. 

To analyze the generation of swirling strength only in the vertical direction, we multiply each term in the swirling strength equation by the vertical component of the axis of rotation vector: $T^i_z = T^i (\mathbf{u}_r)_z$ \citep{canivetecuissa2020}. The same procedure can be applied to compute the horizontal components of the terms in the swirling strength generation equation.

Figure~\ref{fig:ss_generation_terms} shows the spatio-temporal absolute averages of each of the vertical terms given by Eq.~\ref{eq:ss_terms}. We observe that in all of the models, the dominant mechanism for generating swirling strength in the convection zone is the hydrodynamic baroclinic term. At this point, the behavior depends on the magnetic field configuration of the models. In SSD simulations, the magnetic baroclinic term becomes comparable with the hydrodynamic baroclinic term across the solar atmosphere. In contrast, unipolar models show a transition in the photosphere where magnetic baroclinic and magnetic tension become the dominant drivers of swirling strength generation. This transition takes place at a lower atmospheric height in Bz200 due to the stronger magnetic field of the simulation. The contribution of the stretching term is negligible compared with the other terms in all models.

The SSD simulations are characterized by a large plasma beta across the entire domain. However, even in this case, the magnetic pressure plays an important role in generating swirling strength across the solar atmosphere. This indicates that the presence of weak fields can enhance the generation of vortical motions. We observe that the addition of a vertical magnetic field in the simulations modifies the main mechanisms of swirling strength generation. There, magnetic terms are the dominant drivers of vortex generation. We also note that the contribution of the hydrodynamic baroclinic term is stronger in the Bz200 model than in the Bz50 and SSD simulations at the same resolution. Stronger fields can increase the density and pressure gradients in magnetic field concentrations. This effect favors the generation of vortex regions through the hydrodynamic baroclinic term.

We find that a higher spatial resolution increases the magnitude of the production of swirling strength. A finer spatial grid allows for better resolution of small structures and turbulence, leading to a systematic increase of these quantities. We also note that the relative importance of the magnetic pressure term increases in relation to the hydrodynamic baroclinic term from the bottom boundary to the solar surface with spatial resolution. This effect is especially visible in the SSD and Bz50 simulations from $20$ km to $10$ km models. \citet{khomenko2025} show that there is a scaling in magnetic field strength in SSD and Bz50 simulations at sub-surfaces layers, which can explain these results.

The dashed black line of Fig.~\ref{fig:ss_generation_terms} shows the total generation of swirling strength ($\sum_i T^i$). As previously noted by \citet{canivetecuissa2020}, the generation of swirling strength due to both baroclinic terms is higher than the total production of swirling strength. This happens because both terms compensate each other as they have opposite effects. For example, regions with a positive magnetic pressure gradient lead to regions with a negative gas pressure gradient. We have similar results to those of \citet{canivetecuissa2020} for the Bz50 model at $10$ km resolution (which is the model most similar to the one used by these authors). This behavior indicates that the effect of the sum of both baroclinic terms drops in the photosphere and becomes comparable with the magnetic tension term. We found a similar effect in all of the simulations. In addition, the Bz200 model shows a different picture in the upper layers. There, both magnetic terms are above the total swirling strength. This means that these two terms appear to compensate each other. We suggest that the balance of forces between the magnetic tension and the magnetic pressure is more critical in the Bz200 model than in the Bz50 simulation, as gas pressure has a lower contribution due to the stronger fields. We believe that this line of work deserves further study, which is beyond the scope of the present paper.

\subsection{$2$D spatial distribution}\label{sec:2d_spatial}

\begin{figure*}[htbp]
    \centering
    \begin{subfigure}[b]{0.3\textwidth}
        \includegraphics[width=.95\textwidth]{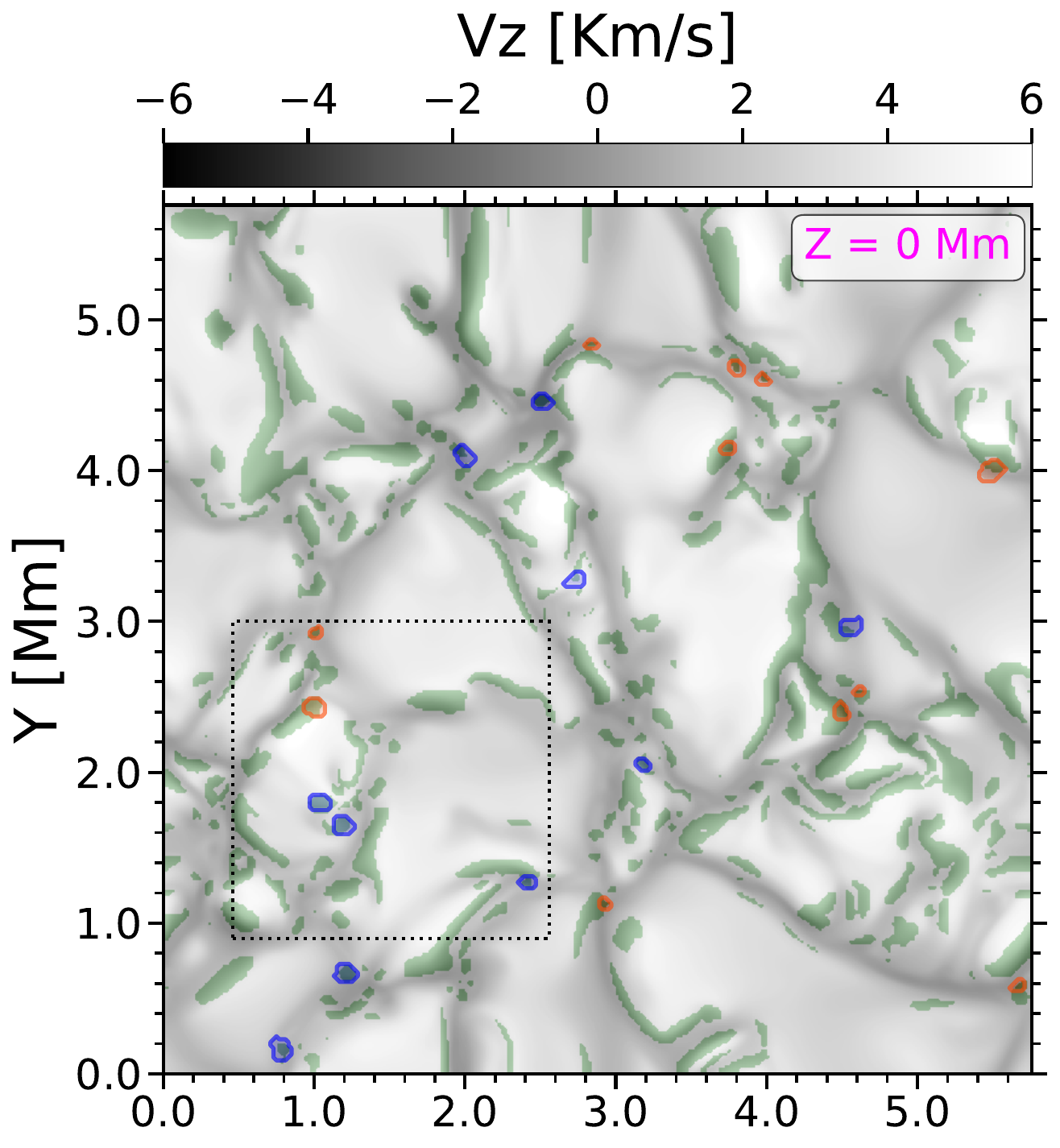}
    \end{subfigure}
    \hfill
    \begin{subfigure}[b]{0.3\textwidth}
        \includegraphics[width=.9\textwidth]{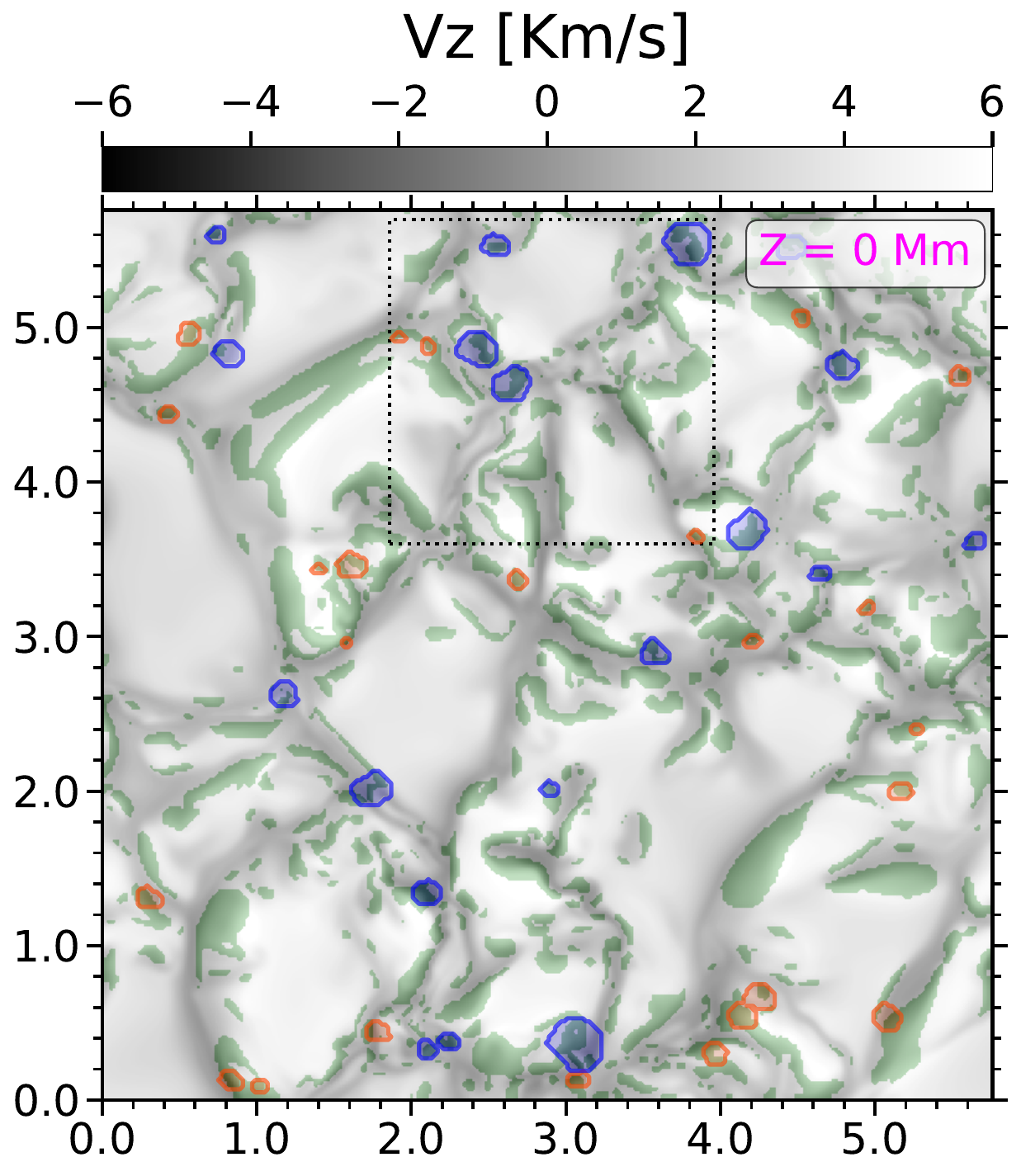}
    \end{subfigure}
    \hfill
    \begin{subfigure}[b]{0.3\textwidth}
        \includegraphics[width=.9\textwidth]{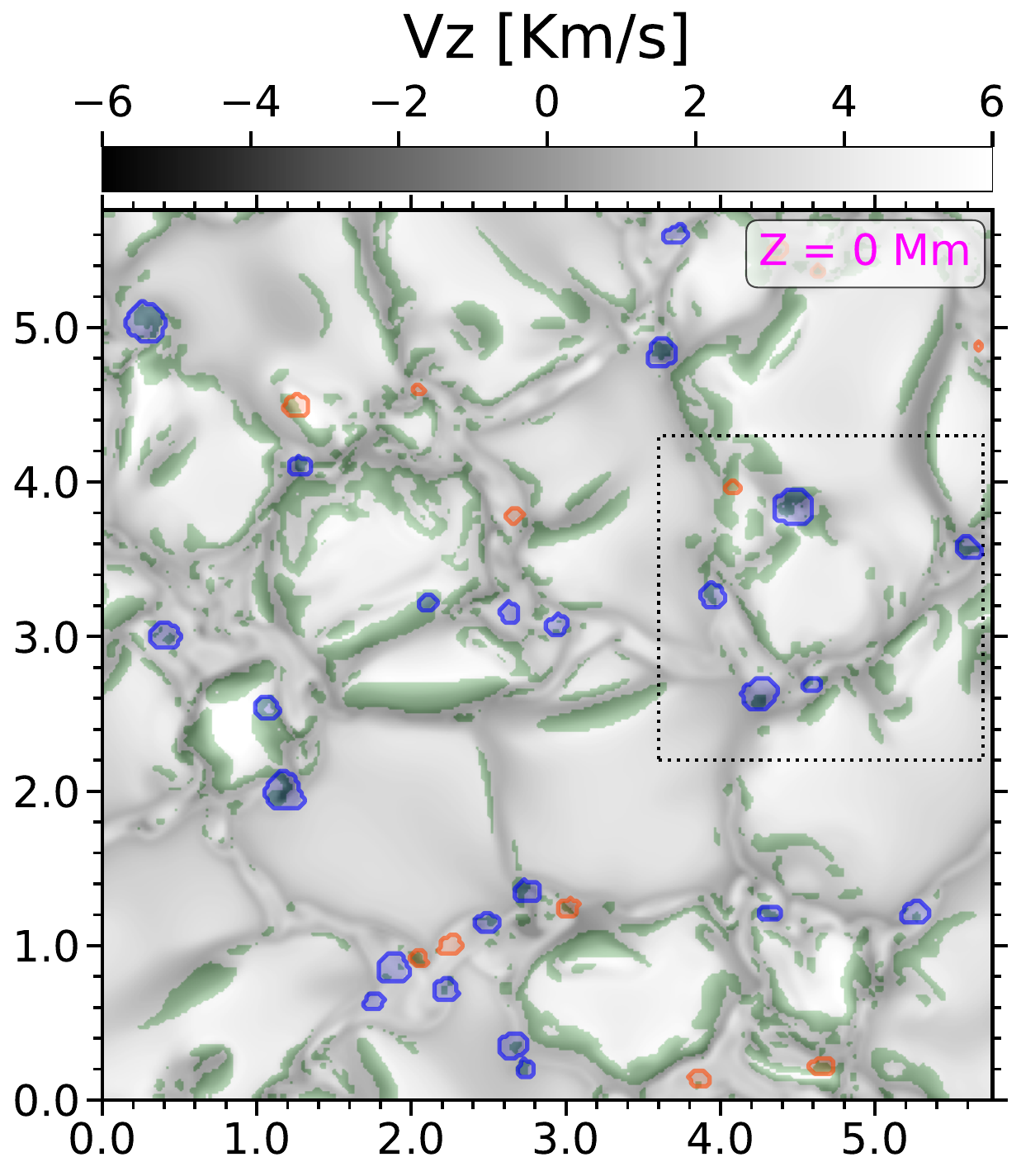}
    \end{subfigure}
    
    \par\bigskip 
    
    \begin{subfigure}[b]{0.3\textwidth}
        \includegraphics[width=.95\textwidth]{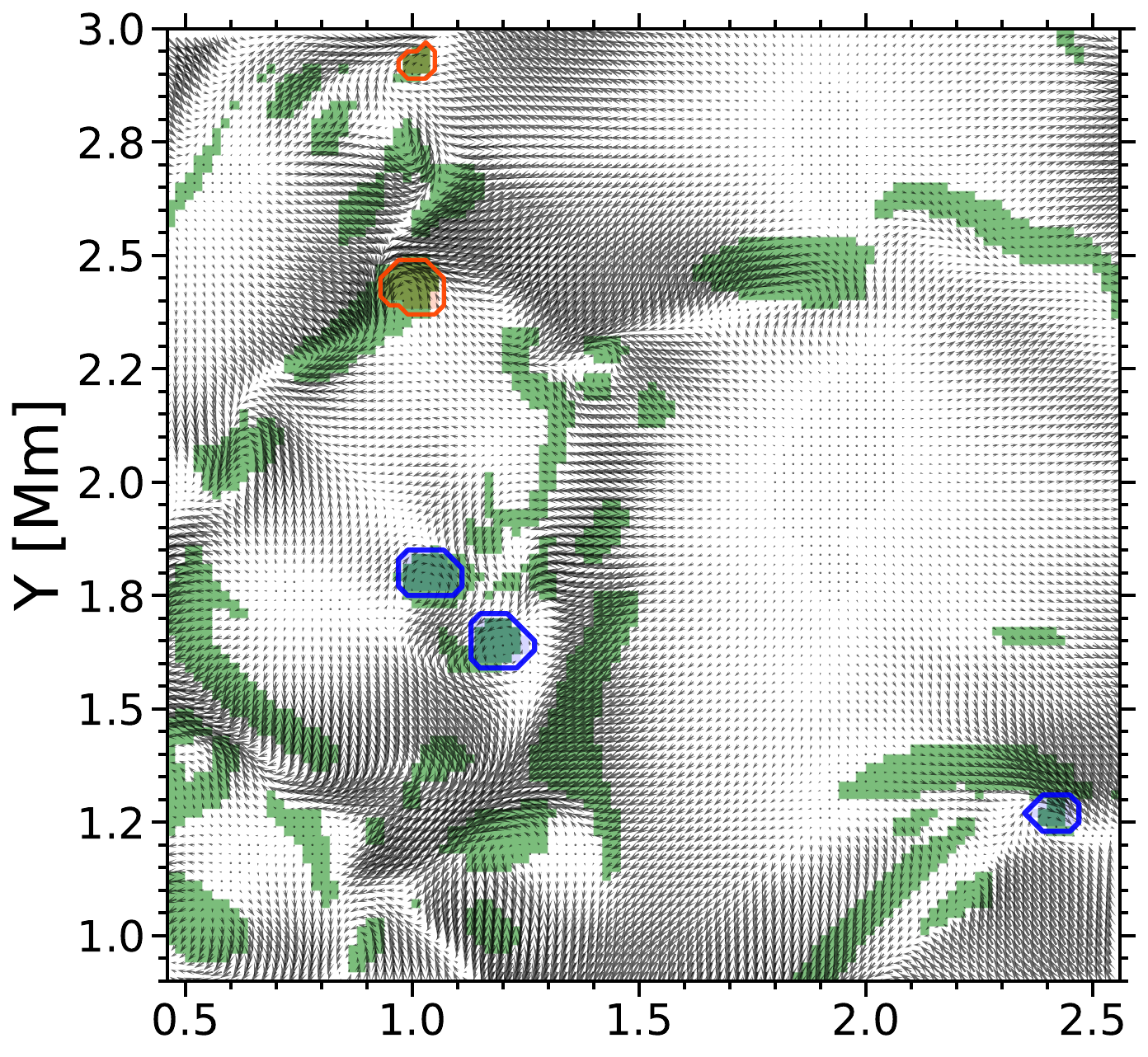}
    \end{subfigure}
    \hfill
    \begin{subfigure}[b]{0.3\textwidth}
        \includegraphics[width=.9\textwidth]{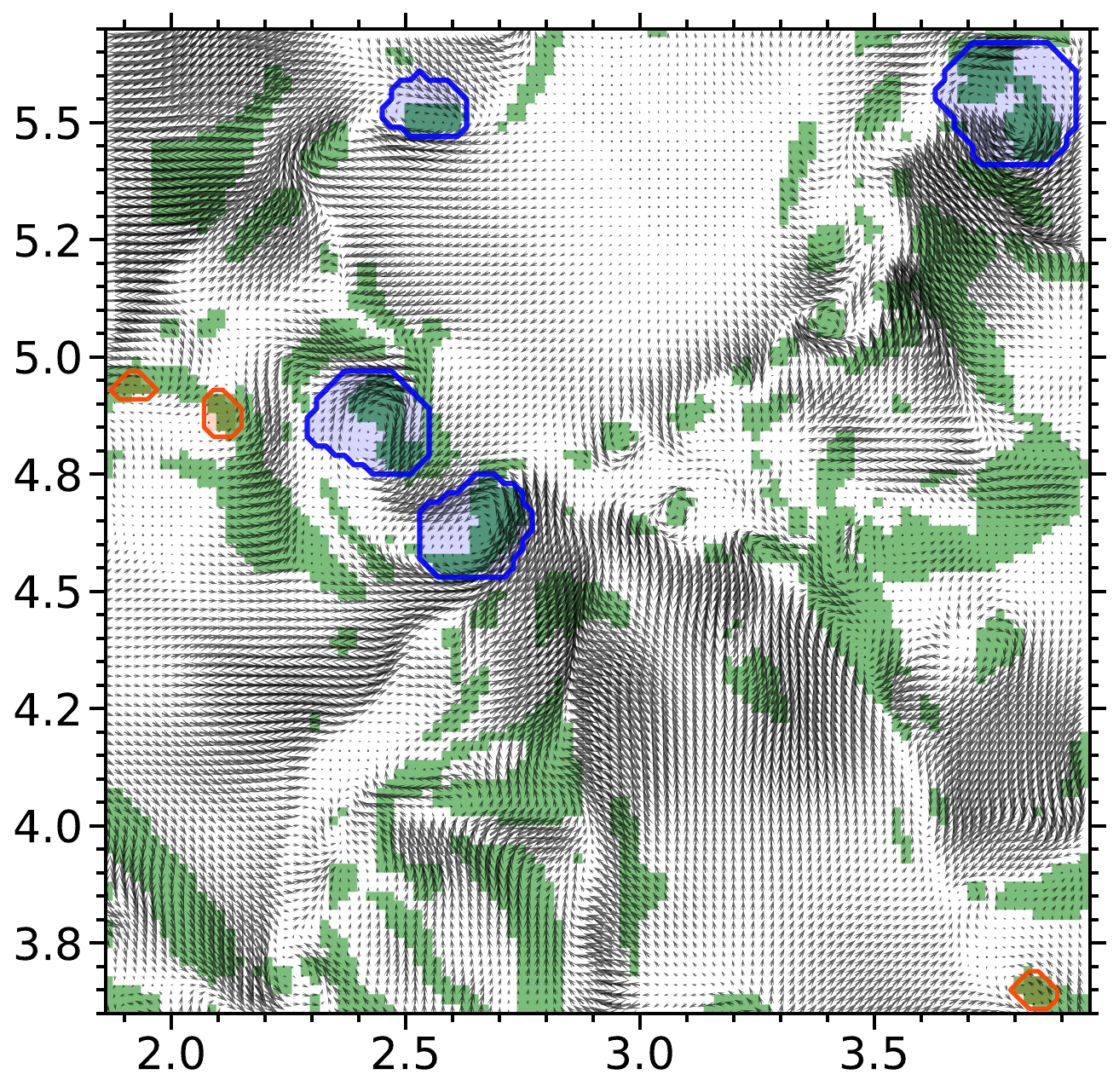}
    \end{subfigure}
    \hfill
    \begin{subfigure}[b]{0.3\textwidth}
        \includegraphics[width=.9\textwidth]{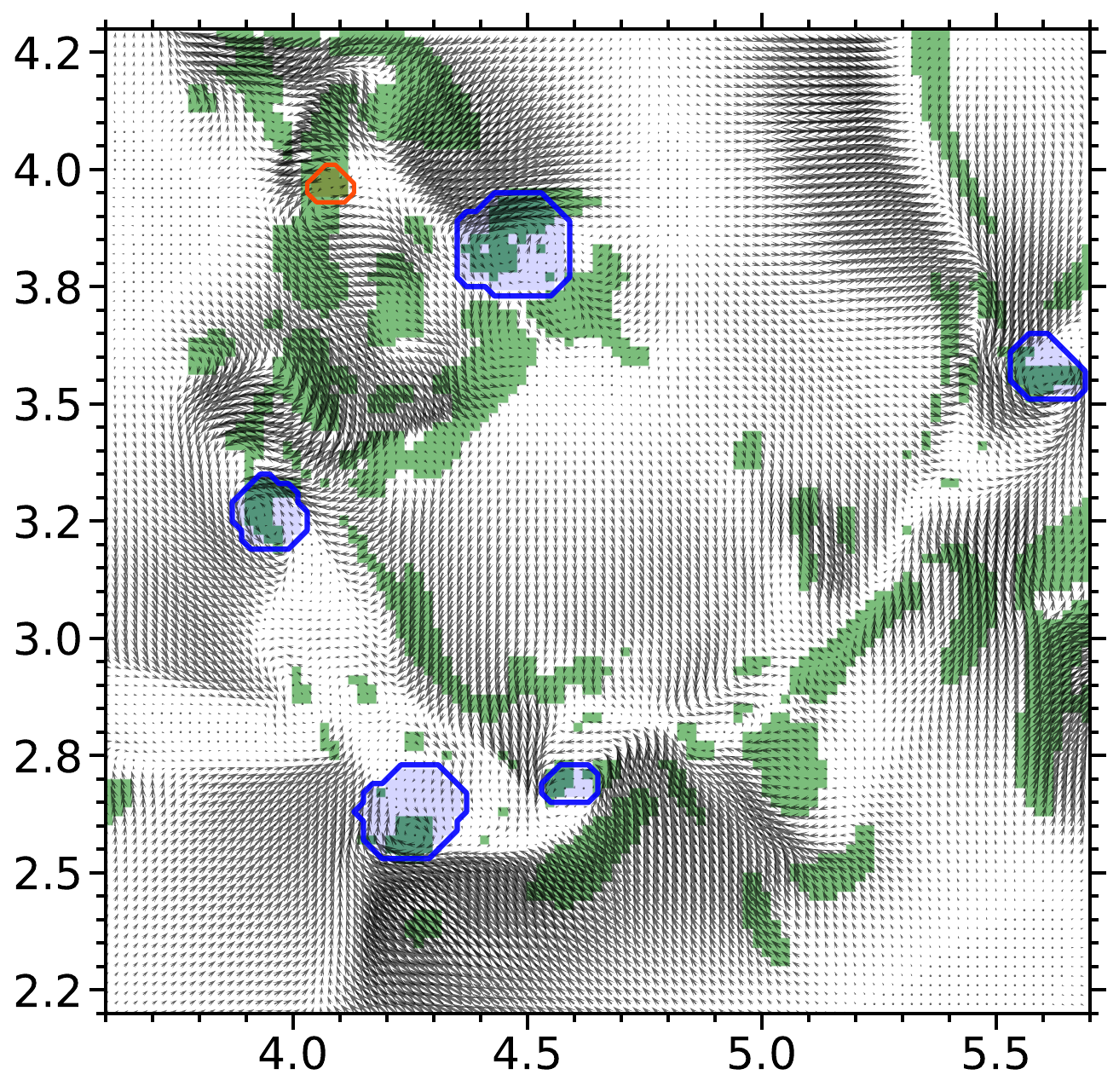}
    \end{subfigure}

    \par\bigskip
    
    \begin{subfigure}[b]{0.3\textwidth}
        \includegraphics[width=.95\textwidth]{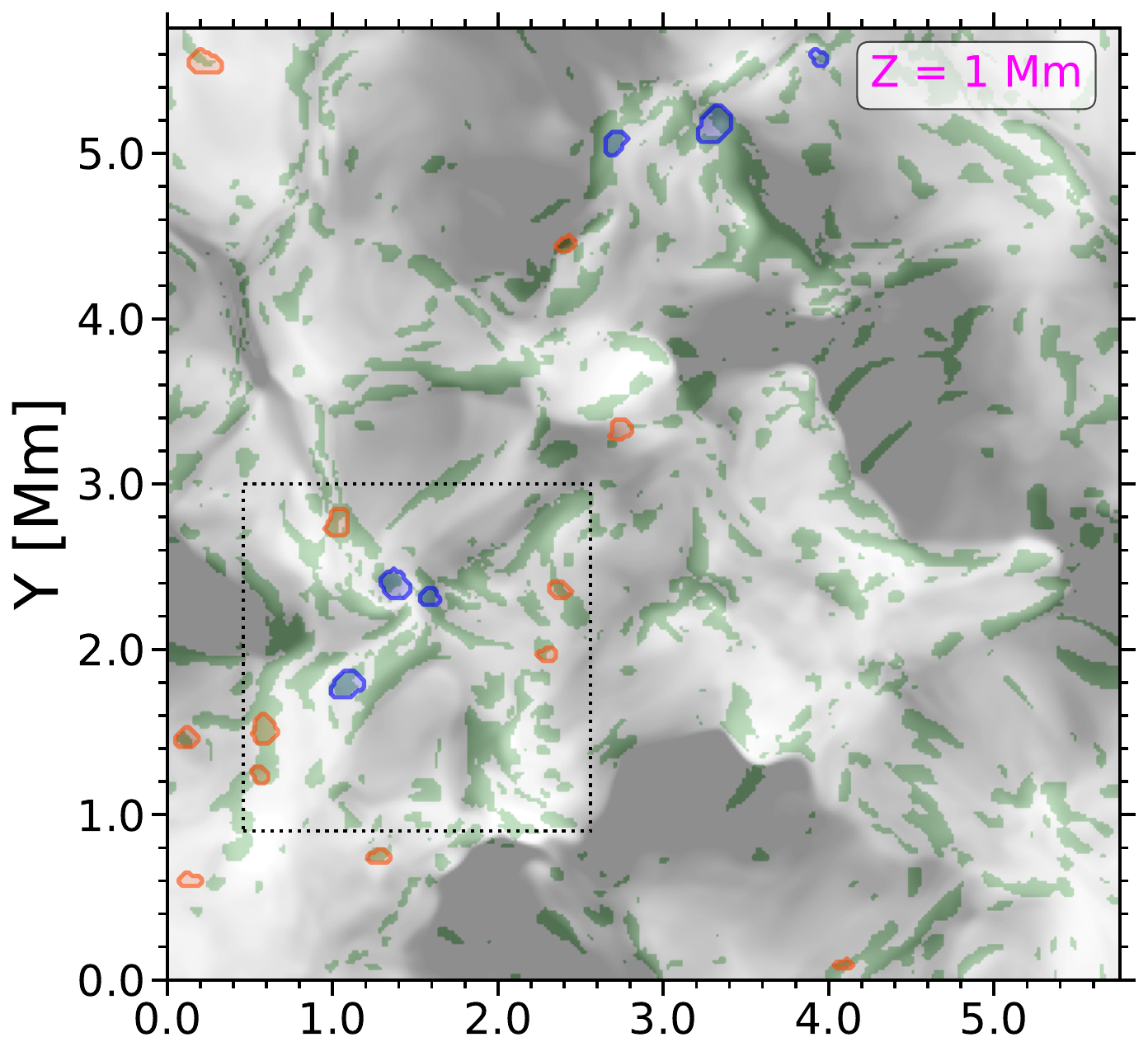}
    \end{subfigure}
    \hfill
    \begin{subfigure}[b]{0.3\textwidth}
        \includegraphics[width=.9\textwidth]{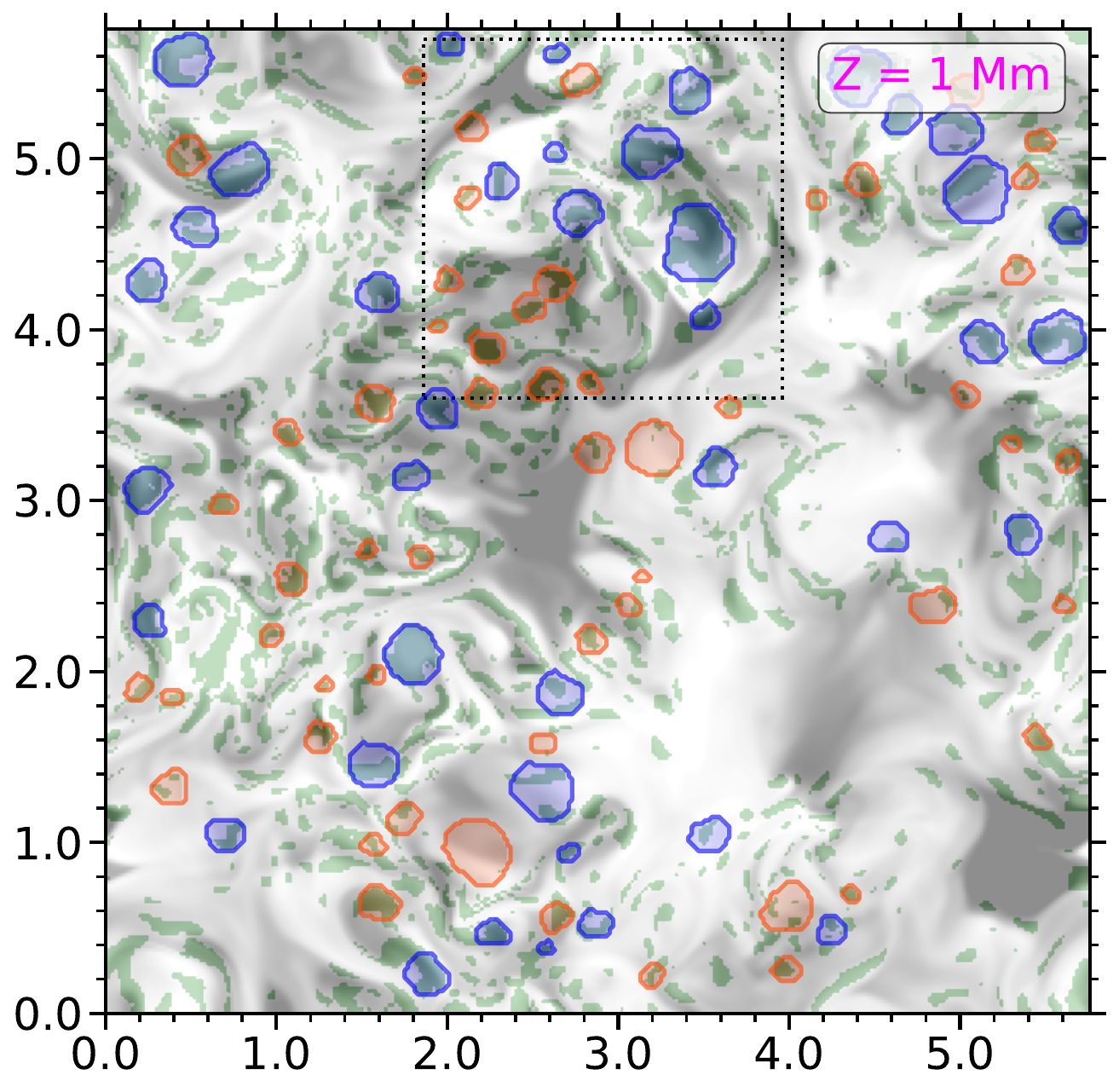}
    \end{subfigure}
    \hfill
    \begin{subfigure}[b]{0.3\textwidth}
        \includegraphics[width=.9\textwidth]{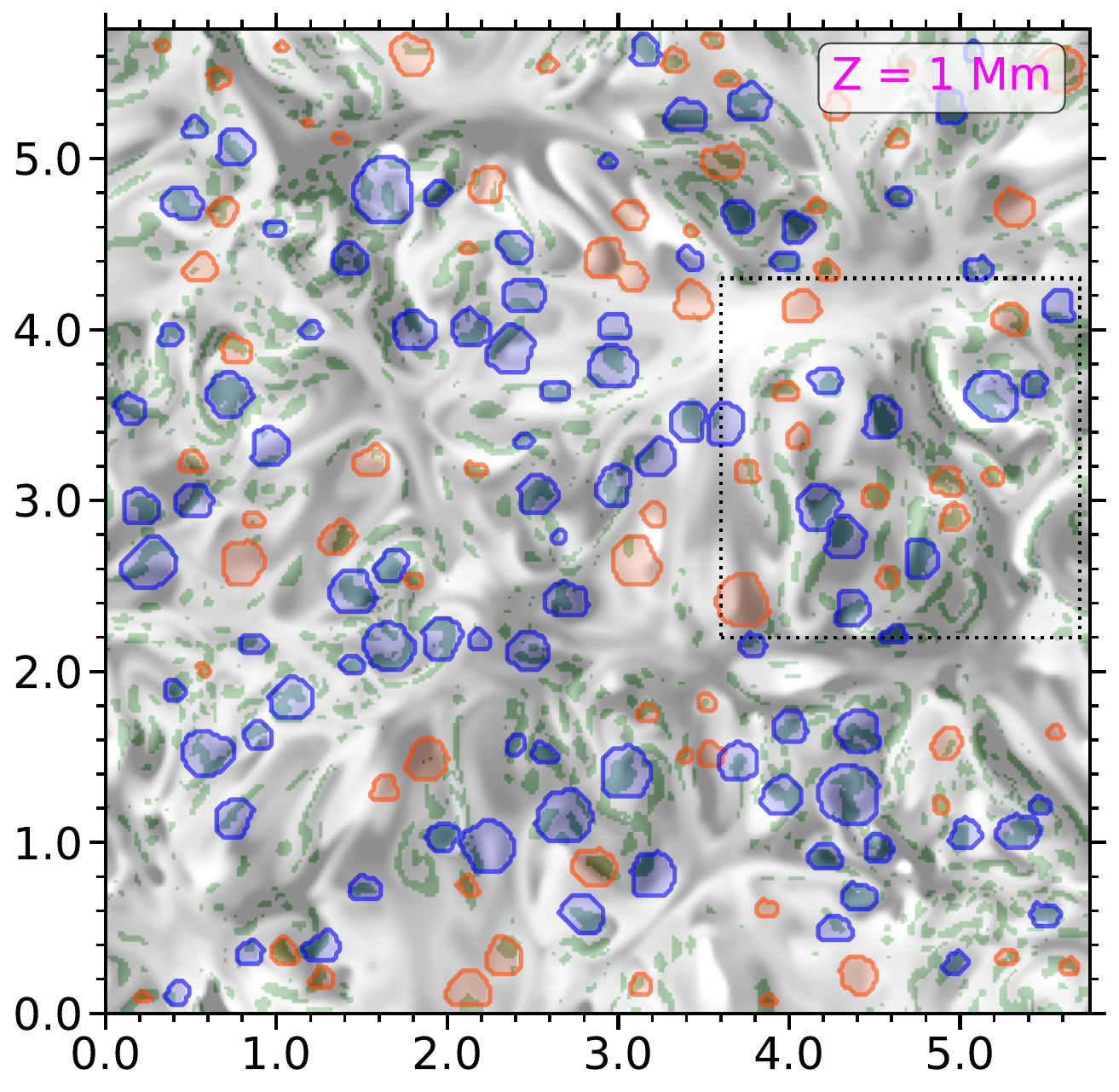}
    \end{subfigure}
    
    \par\bigskip
    
    \begin{subfigure}[b]{0.3\textwidth}
        \includegraphics[width=.95\textwidth]{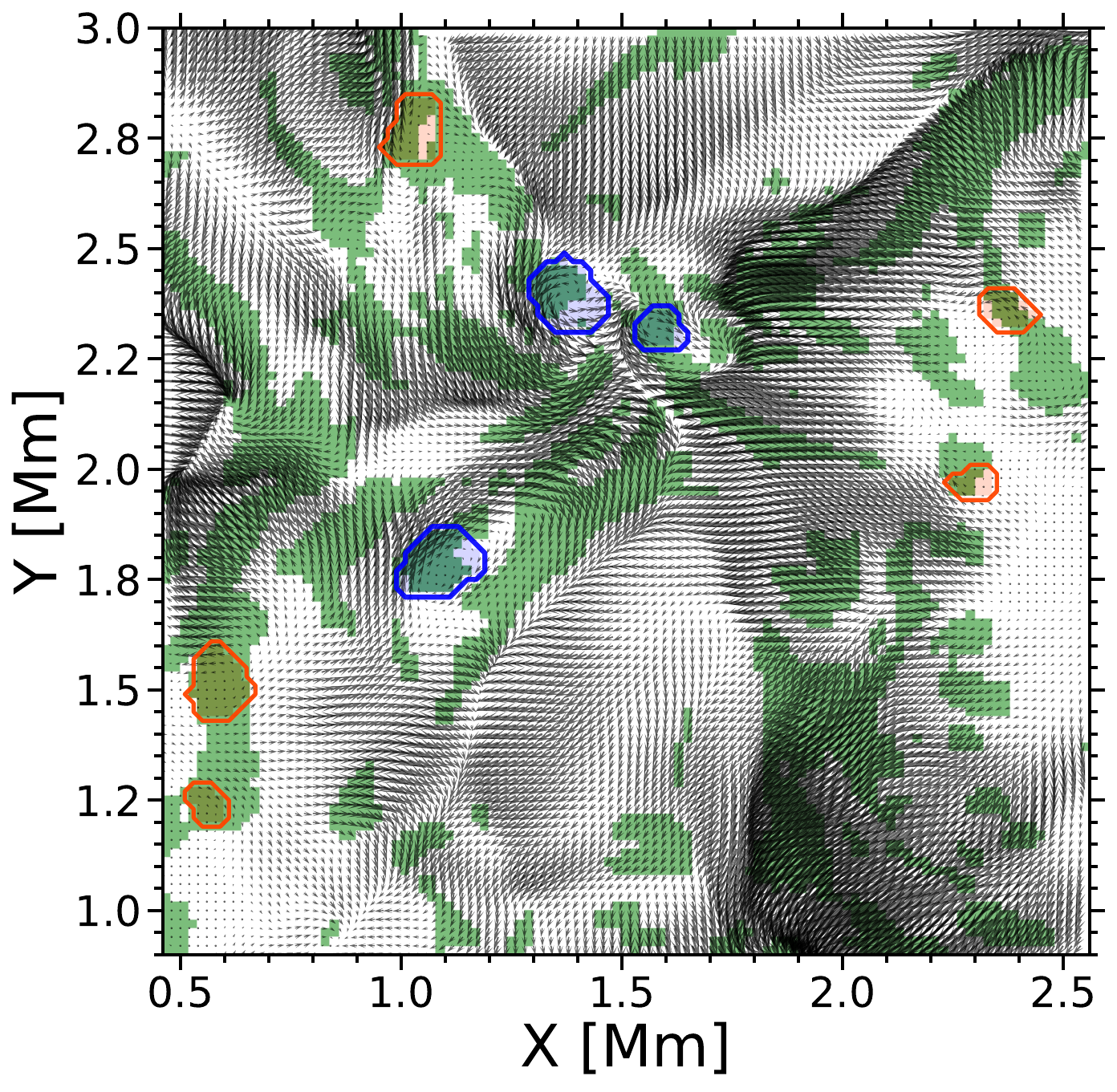}
    \end{subfigure}
    \hfill
    \begin{subfigure}[b]{0.3\textwidth}
        \includegraphics[width=.9\textwidth]{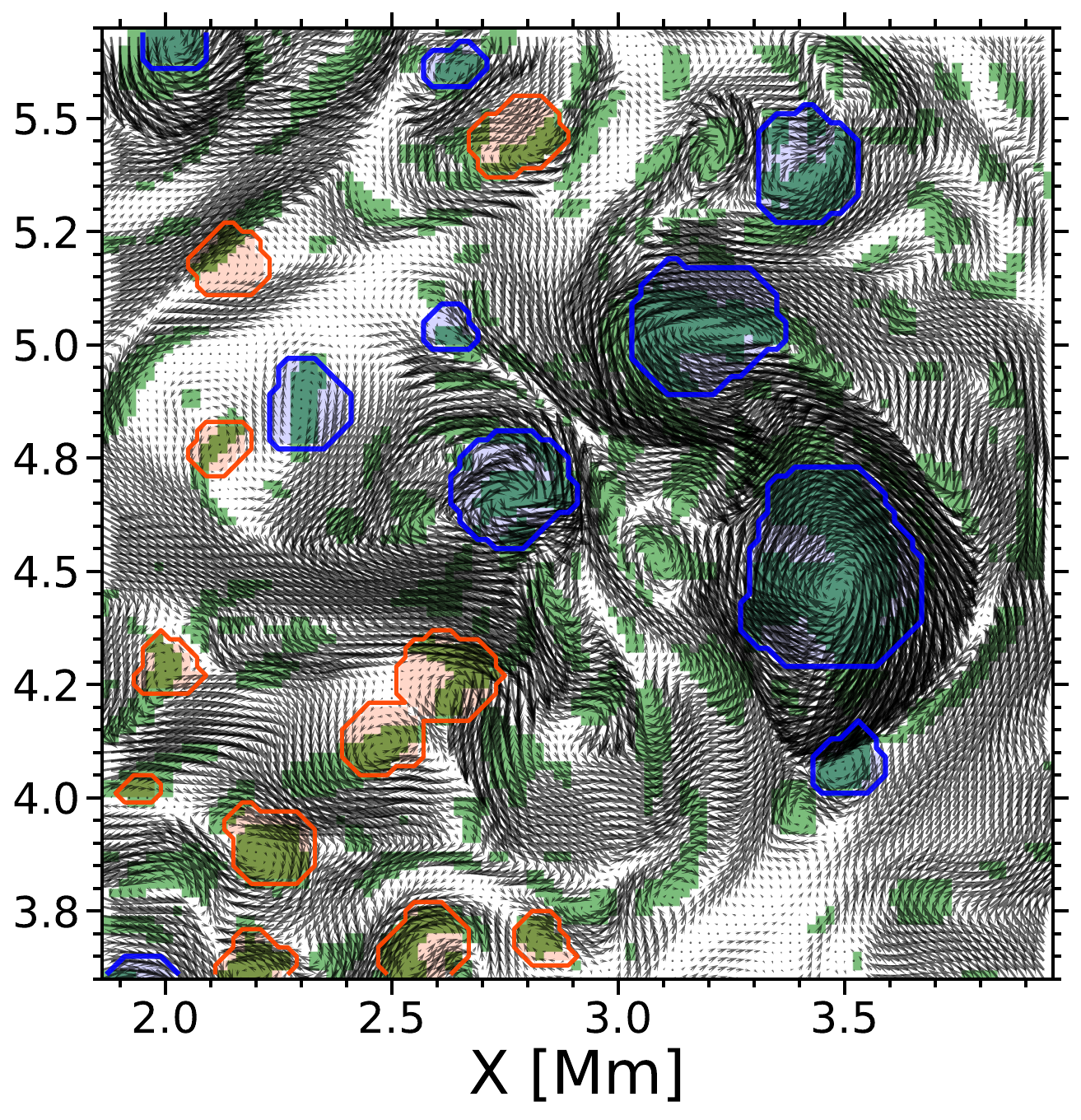}
    \end{subfigure}
    \hfill
    \begin{subfigure}[b]{0.3\textwidth}
        \includegraphics[width=.9\textwidth]{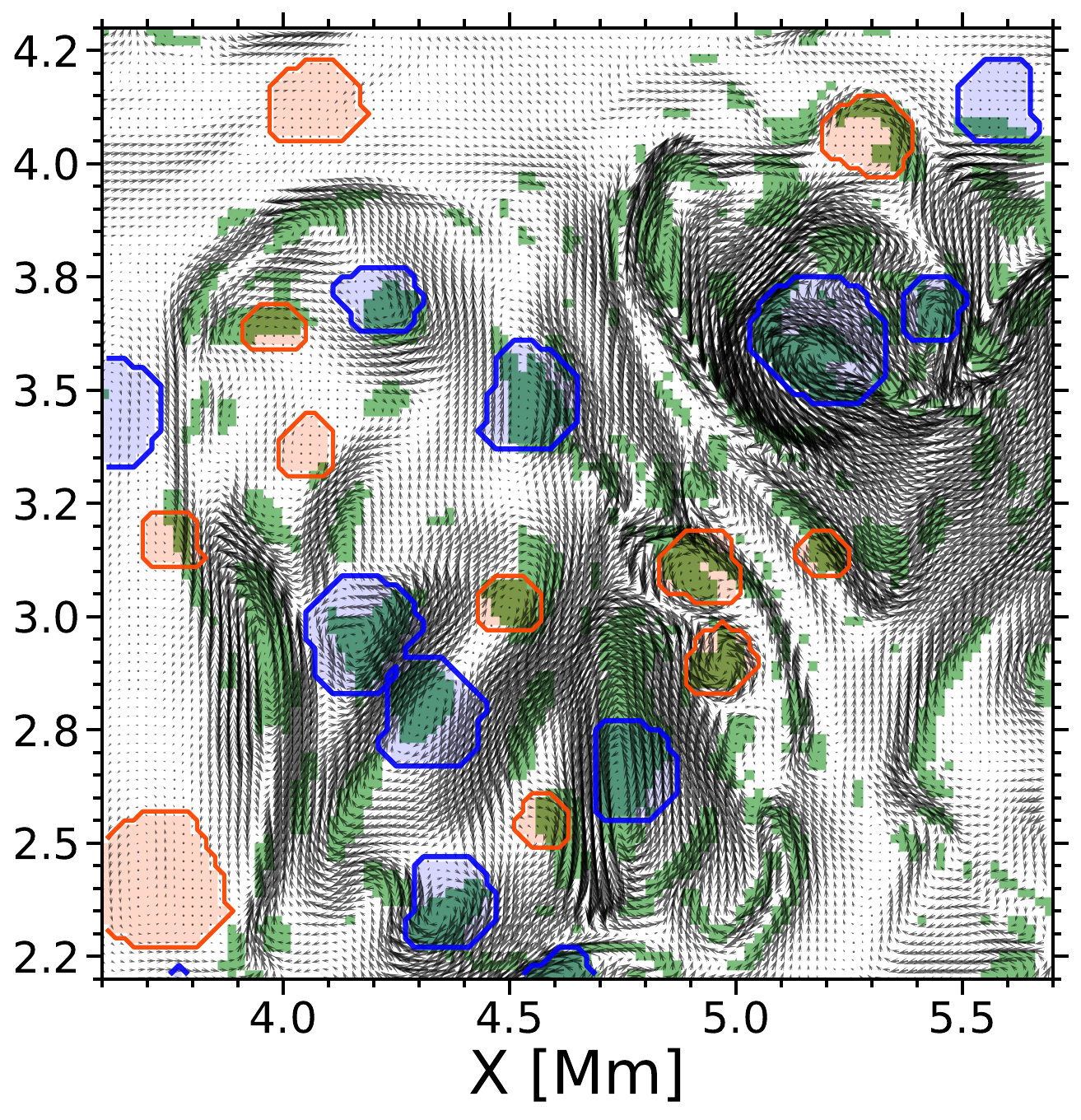}
    \end{subfigure}
    
    \caption{Spatial distribution of vortex detections in the photosphere (first and second rows) and in the chromosphere (third and fourth rows) obtained with swirling strength (green patches) and the SWIRL code (orange and blue contours). From left to right, columns show results of SSD, Bz50 and Bz200 simulations at $20$ km resolution. First and third rows shows the global domain at the photosphere and chromosphere, respectively. Second and fourth rows show a close-up view given by the black dotted square in the first and third rows, respectively. Black arrows in the zoom plots represent the horizontal velocity field ($v_x$, $v_y$). The arrows scale is the same in all the panels.}
    \label{fig:spatial_distribution}
\end{figure*}

Figure \ref{fig:spatial_distribution} shows a comparison of vortex detections produced by swirling strength (green patches) and the SWIRL code (orange and blue contours) at two different heights. From left to right, each column corresponds to the SSD, Bz50, and Bz200 models at $20$ km horizontal spatial resolution. As discussed in Sec.~\ref{sec:swirling_strength}, the swirling strength method requires a threshold to discard weak vortices and noisy regions, which depends on the model and the spatial resolution of the simulation. Thus, Fig.~\ref{fig:spatial_distribution} only shows swirling strength regions which are above the height-dependent criterion (given by Fig.~\ref{fig:v_rms_and_ss_var_criterion}). The orange contours from SWIRL represent the raw detections after only applying the convex hull algorithm, while blue contours from SWIRL represent those vortices that additionally satisfy our complete filter selection (Sec.~\ref{sec:SWIRL}). 

At $Z=0$ Mm (first and second row), the regions detected by swirling strength are mainly distributed in elongated structures in all three models. Comparing these regions with the background vertical velocity map, we observe that swirling strength detections are located along intergranular lanes. This is a consequence of granulation. Upward and downward motions at the edges of granules and intergranules lead to a rotation along a horizontal axis \citep[also reported by][]{moll2011, moll2012}. Appendix \ref{app:horizontal_vortices} shows some examples of vertical slices where the horizontal rotation is illustrated. These results are independent of the magnetic field configuration. All three models exhibit a similar spatial distribution of swirling strength as the underlying physics is hydrodynamic in nature. These large horizontal structures have also been reported in observations as bright and dark edges along intergranular lanes \citep{steiner2010}.

The second row of Fig.~\ref{fig:spatial_distribution} shows a close-up view of the black dotted square areas for each model together with the horizontal velocity field (black arrows). A visual examination reveals numerous locations where rotation around the vertical axis is found. The detected structures cover a range of sizes and exhibit irregular shapes. For instance, in the Bz50 model (middle column) a small vortex can be observed at $(x, y) \sim (3.1 \, \text{Mm}, 4.7 \, \text{Mm})$ and a larger one at $(x, y) \sim (2.7 \, \text{Mm}, 4.6 \, \text{Mm})$. We notice that parts of the large vortices remain undetected by swirling strength as they  often exhibit smoother velocity gradients in certain regions. Consequently, the swirling strength magnitude decreases and may fall below the detection threshold, resulting in a partial loss of the vortex structure. Smaller vortices are not affected by this phenomenon.

We also note small regions detected by swirling strength composed of only a few grid points. Visually, they are not associated with a rotation of the horizontal velocity field. These features can be considered as noise of the method. They exceed the threshold but are not associated with vortex flows. 

Vortices detected by the SWIRL code exhibit a different morphology compared to the results of swirling strength. These structures are also located in intergranular lanes but are significantly more compact, having an almost circular shape. The detections have visibly better correspondence to the horizontal velocity field, as SWIRL relies exclusively on these components (Sec.~\ref{sec:SWIRL}). Therefore, this method enhances the identification of vertically oriented vortices, being not able to detect horizontal vortex structures. 

SWIRL attempts to detect regions with a closed rotating velocity field through its clustering algorithm. Thus, it can account for whether the surroundings of a given grid point are also rotating. This results in compact and isolated structures. Now, the large vortex at $(x, y) \sim (2.7 \, \text{Mm}, 4.6 \, \text{Mm})$ in the Bz50 model, is perfectly matched to the rotating velocity field. Most of the SWIRL vortices in the photosphere are accurately represented. 

The filtered-out vortices (orange contours) are generally the smallest structures or those with lower rotation speeds. Sometimes, properly detected vortices are filtered out of our selection. This arises mainly from the temporal filtering process, meaning that these structures are short-lived. An example of such scenario is the vortex in the SSD model around $(x, y) \sim (1.0 \, \text{Mm}, 2.4 \, \text{Mm})$ at $Z=0$ Mm (second row, first column), although it is an uncommon situation. 

Unipolar models yield a significantly higher number of detections when analyzed with the SWIRL code compared to the SSD simulations. This indicates that the magnetic field favors the generation of vertically oriented vortices. As we noted in Sec.~\ref{sec:ss_gen_terms}, stronger fields enhance the generation of swirling strength through the hydrodynamic and magnetic baroclinic terms. As a consequence, a larger number of vortices are found. Previous studies have analyzed the vorticity equation in strong magnetic field simulations \citep{shelyag2011, silva2024b} and concluded that the primary mechanism for vortex generation is the magnetic baroclinic term. However, these results should be taken with caution, as the vorticity equation may not be the most suitable tool for this purpose \citep{canivetecuissa2020}. In addition, the sizes of the detections are also affected by the magnetic field. We notice a slight increase in the vortex area in unipolar models with respect to SSD simulations.

The third and fourth rows of Fig.~\ref{fig:spatial_distribution} show the results at $Z=1$~Mm at the same time instant. As we move into higher layers of the solar atmosphere, the turbulence of the plasma increases, as can be seen in the fourth row. Therefore, the performance of the methods may be affected due to the complexity of the flows. 

In SSD simulations, regions detected by swirling strength also exhibit an elongated shape, as at $Z=0$~Mm, but now located along shock fronts. \cite{moll2012} and \cite{battaglia2021} reported similar results in a non-magnetic simulation. These detections are believed to be a consequence of the transit of shock fronts, which can generate horizontal vortices due to the strong pressure gradients through the baroclinic term (see Appendix \ref{app:horizontal_vortices}). We also identify a few vertically oriented vortices at these heights, but their number is scarce.

The picture in simulations with a unipolar vertical field changes significantly with respect to SSD models. Plasma dynamics is no longer dominated by the interference pattern of shock waves. Instead, a more structured atmosphere develops due to the presence of the magnetic field. Thus, swirling strength detections are more compact and localized. From the horizontal velocity field, it can be clearly observed an increase in the number of vertically oriented vortices. This is a consequence of the vertical magnetic field of the simulations, since plasma tends to organize around it. Swirling strength can properly capture those regions, as e.g. vortices at $(x, y) \sim (3.2 \, \text{Mm}, 5.4 \, \text{Mm})$ or at $(x, y) \sim (3.5 \, \text{Mm}, 4.5 \, \text{Mm})$ in the Bz50 simulation. Vertical magnetic field suppresses the formation of tilted or horizontal vortices. Here, we note larger sizes than in the solar photosphere.

The performance of SWIRL in the chromosphere produces accurate results. Detections in SSD and unipolar models exhibit a good agreement with the horizontal velocity field (fourth row of Fig.~\ref{fig:spatial_distribution}). The number of detections in SSD simulations is relatively low. This is because the number of vertical vortices is small, being most of vortices horizontally oriented as observed with the swirling strength. In unipolar models we observe an increased number of SWIRL detections. The Bz200 model shows a larger number of vortices compared with the Bz50 simulation at the same height. This suggests that stronger vertical magnetic fields at the same spatial resolution enhance the formation of vertical vortices.

A wide variety of vortex sizes can be identified. As magnetic field lines open with height, plasma rotating around them will also increase their size. Consequently, larger vortices can be developed in simulations with vertical magnetic fields. We note that detections in the Bz50 model have slightly larger sizes than those of the Bz200 simulation, despite having a stronger field. Due to the turbulent plasma conditions, we note that some vortices could be underestimated in size. In the chromosphere, we found more filtered-out vortices than in the photosphere mainly due to their low temporal coherence, such as the vortices at $(x,y) \sim (2.2 \, \text{Mm}, 3.9 \, \text{Mm})$ or at $(x,y) \sim (2.5 \,\text{Mm}, 4.1 \,\text{Mm})$ in the Bz50 model (fourth row, second column from Fig. \ref{fig:spatial_distribution}). In addition, some structures were not detected by SWIRL as the vortices at $(x,y) \sim (3.2 \, \text{Mm}, 5.4 \, \text{Mm})$ and $(x,y)\sim (3.1 \, \text{Mm}, 4.5 \, \text{Mm})$. This effect arises in regions with a high density of vortices, which prevents SWIRL from distinguishing adjacent structures, resulting in missed detections. By adjusting the parameters of SWIRL (see Appendix \ref{app:SWIRL_params}), we determined the best compromise between maximizing the number of detected structures and maintaining the accuracy of the results. Nevertheless, we observe that SWIRL detections are satisfactory when compared to the velocity field.

\subsection{3D vortex structures}\label{sec:3d_vortices}

\begin{figure*}[h!]
    \centering
    % Fila 1
    \begin{subfigure}[b]{\columnwidth}
        \includegraphics[width=.85\columnwidth]{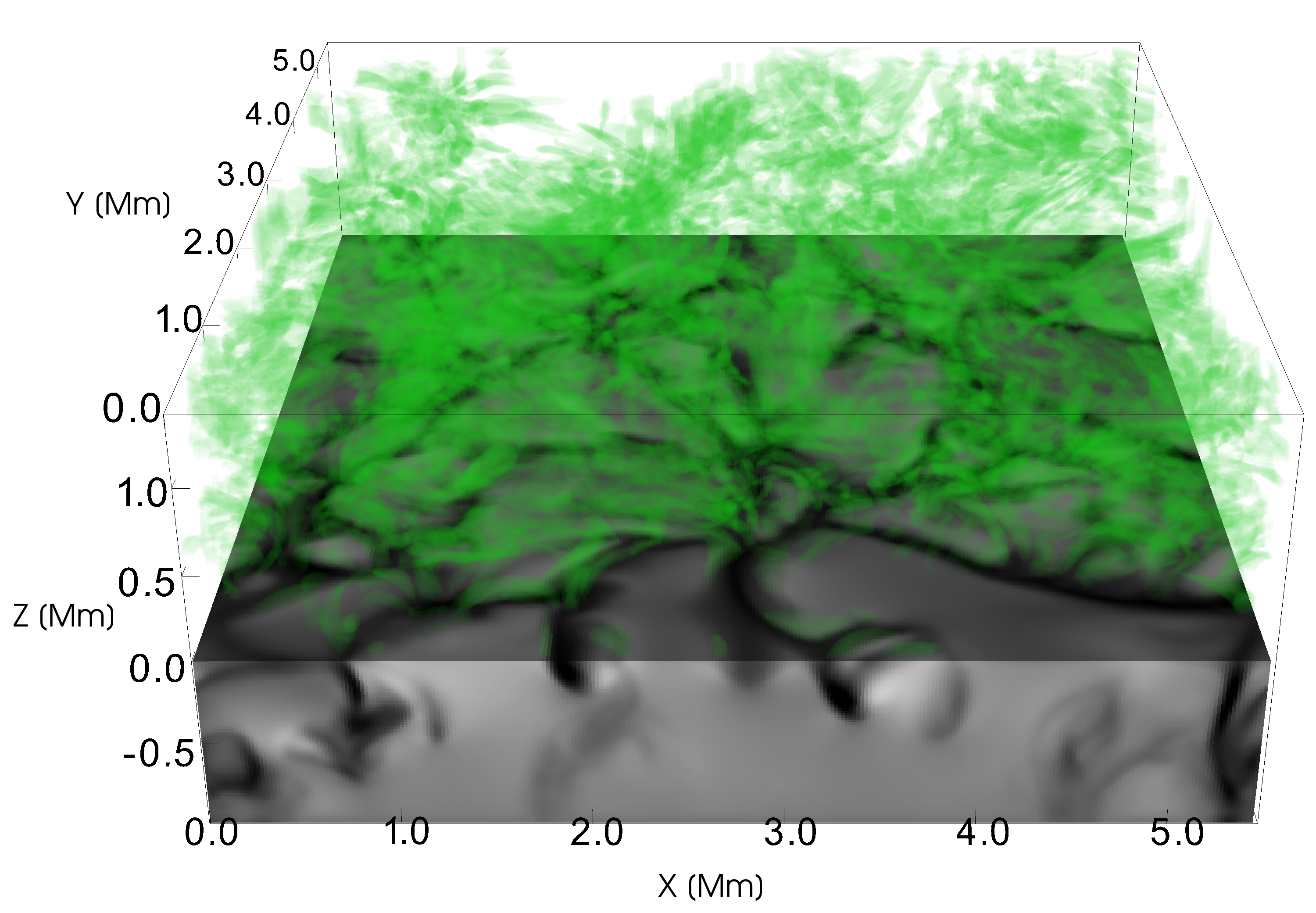}
    \end{subfigure}
    \hfill
    \begin{subfigure}[b]{\columnwidth}
        \includegraphics[width=.85\columnwidth]{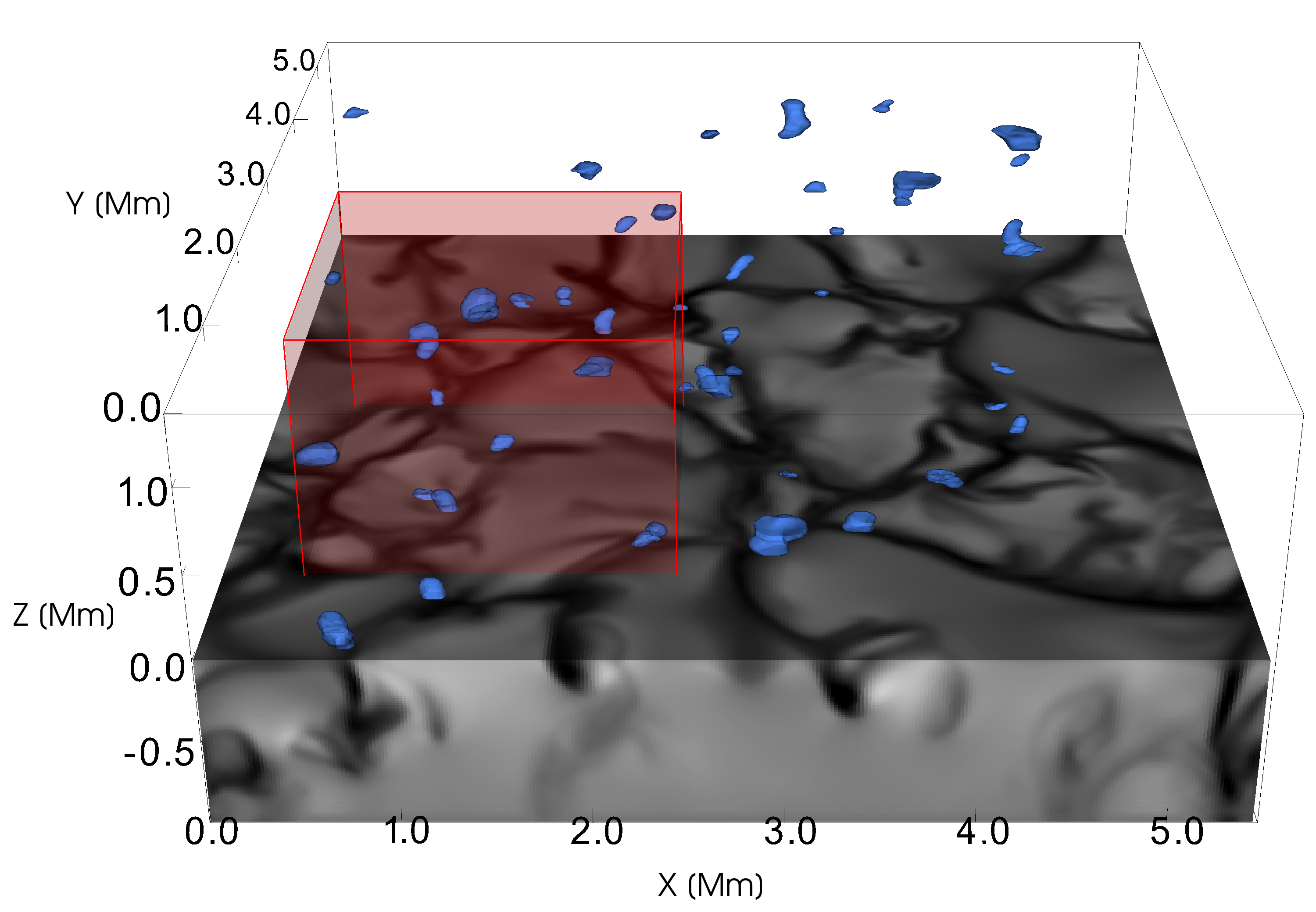}
    \end{subfigure}

    \par\bigskip 

    \begin{subfigure}[b]{\columnwidth}
        \includegraphics[width=.85\columnwidth]{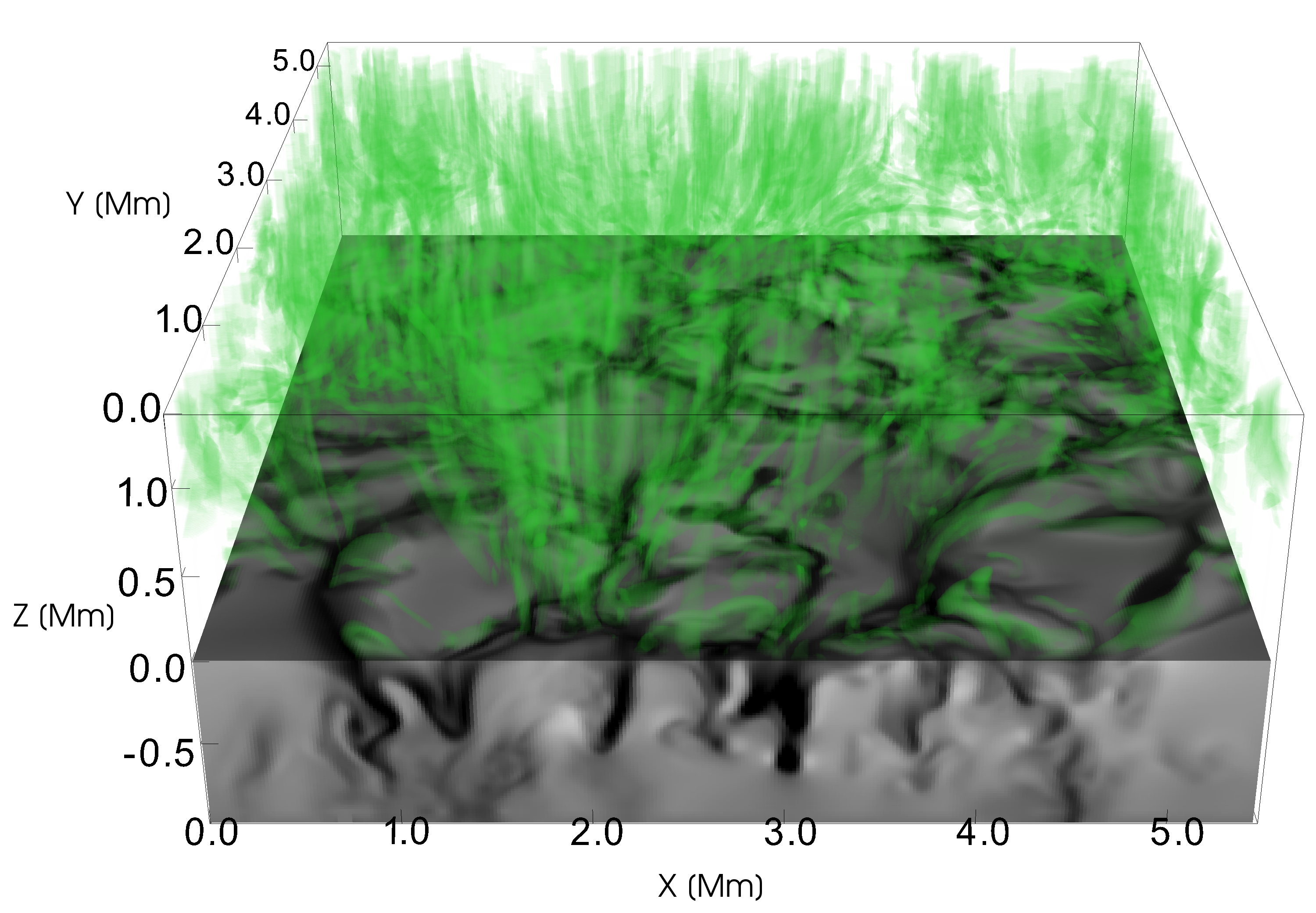}
    \end{subfigure}
    \hfill
    \begin{subfigure}[b]{\columnwidth}
        \includegraphics[width=.85\columnwidth]{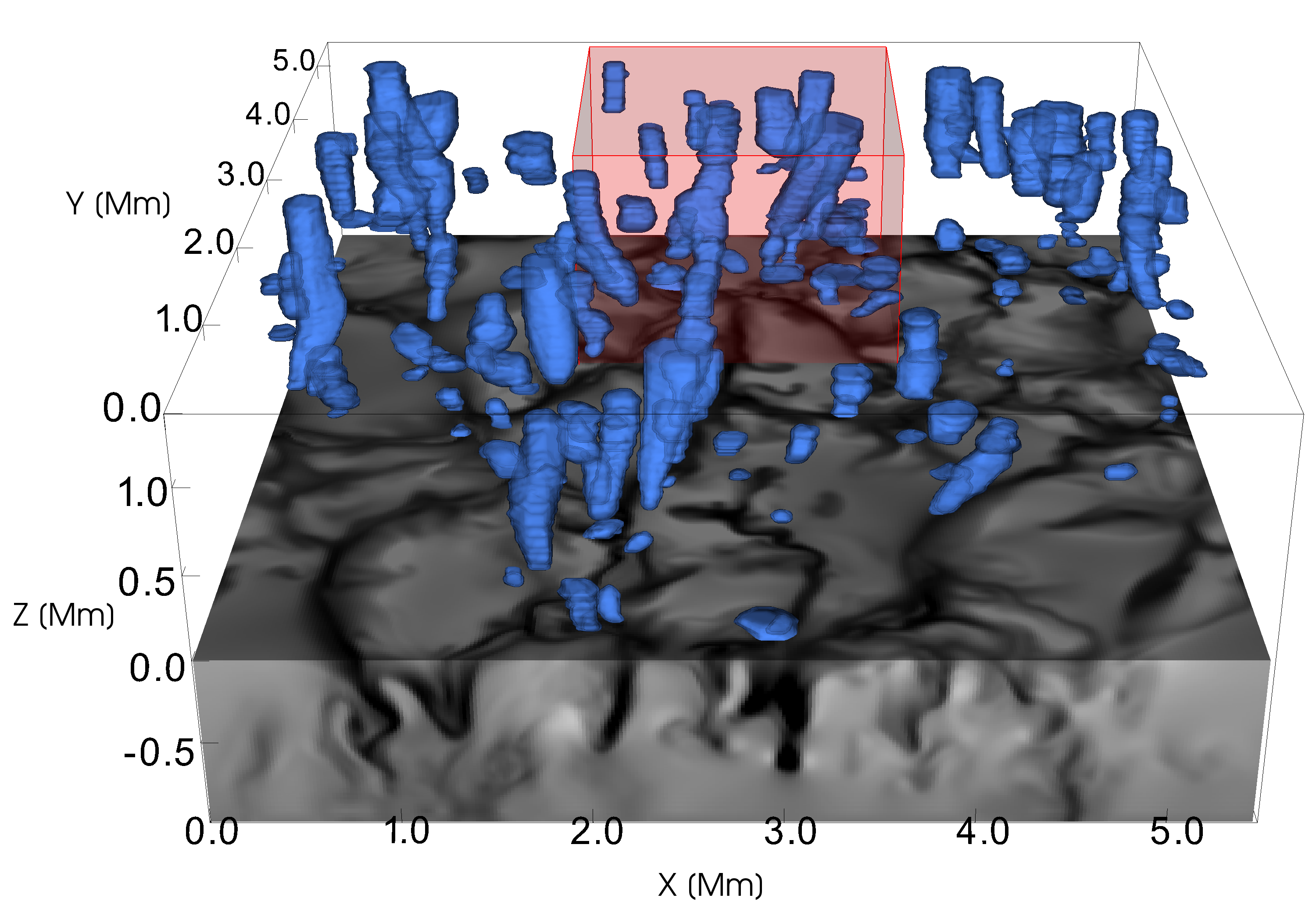}
    \end{subfigure}

    \par\bigskip 

    \begin{subfigure}[b]{\columnwidth}
        \includegraphics[width=.85\columnwidth]{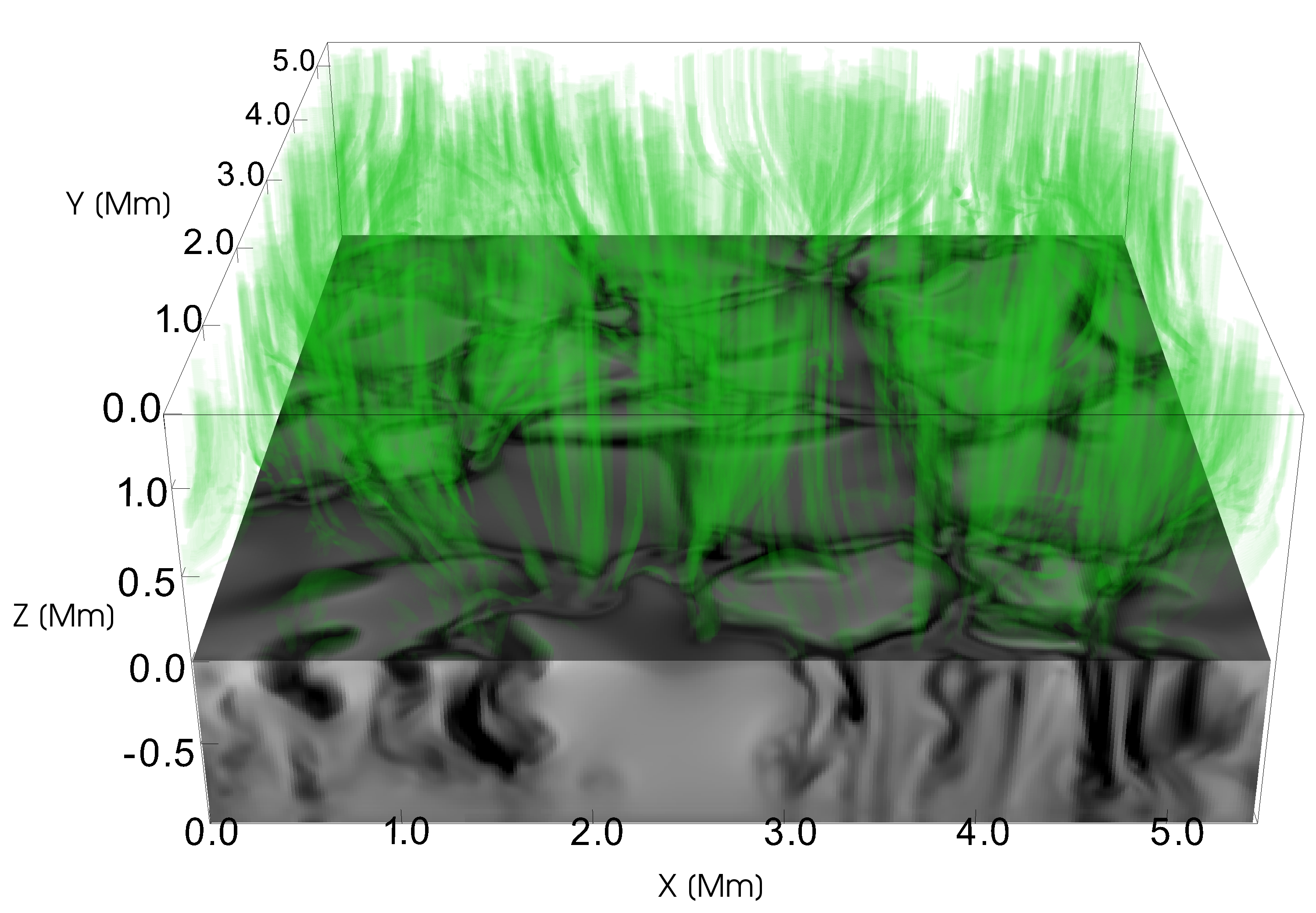}
    \end{subfigure}
    \hfill
    \begin{subfigure}[b]{\columnwidth}
        \includegraphics[width=.85\columnwidth]{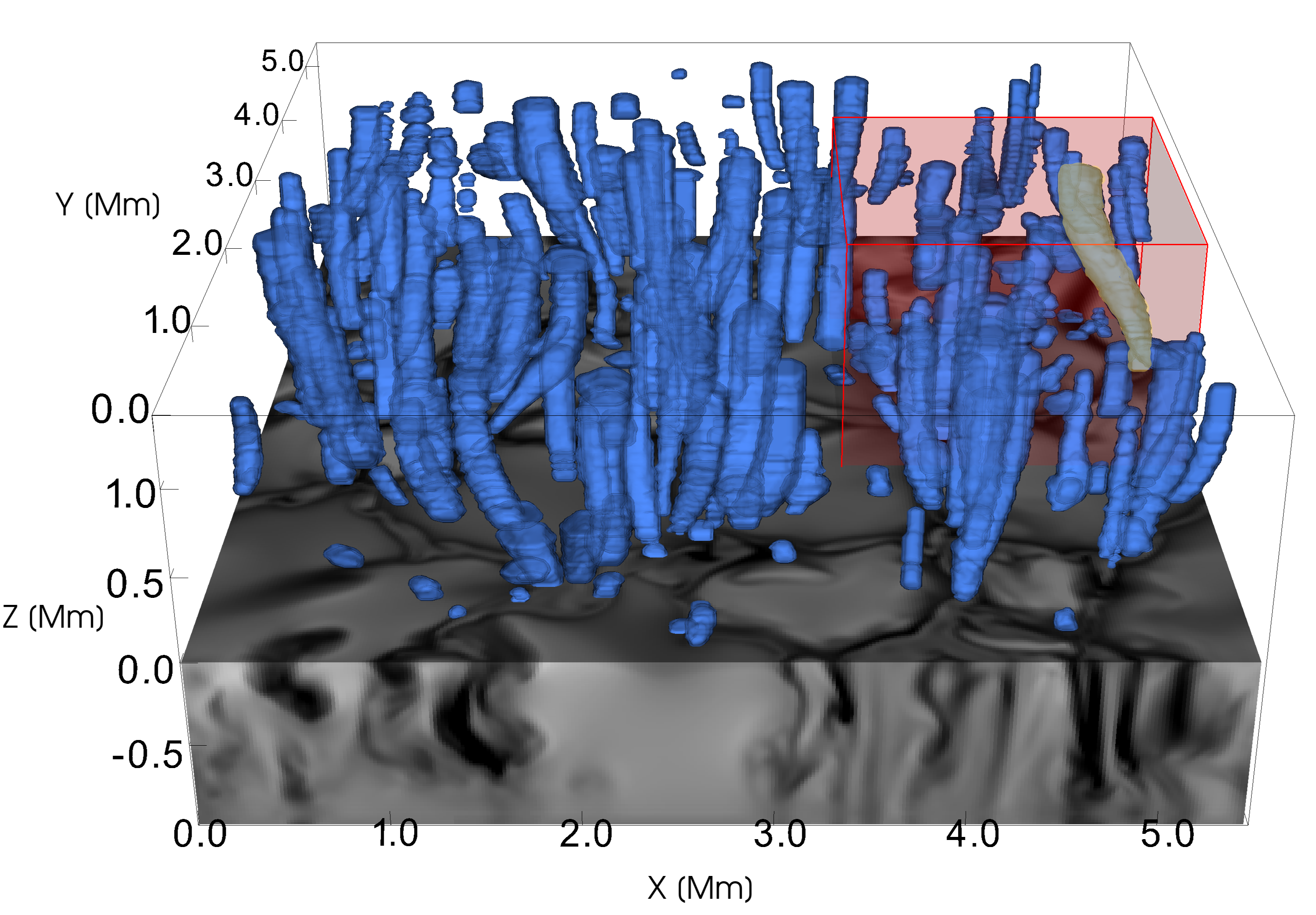}
    \end{subfigure}
    \caption{Three-dimensional renders of vortex detections computed with swirling strength (left column) and SWIRL (right column). Top, middle and bottom panels contain data from SSD, Bz50 and Bz200 simulations, respectively, at the $20$ km spatial resolution. The red cubes indicate the location of the zoomed-in regions in Fig.~\ref{fig:spatial_distribution}. The vertical velocity field is plotted from the bottom of the domain up to $Z=0$ Mm. The color scale is the same as in Fig.~\ref{fig:spatial_distribution}.}
    \label{fig:3d_vortices}
\end{figure*}

Fig.~\ref{fig:3d_vortices} displays three-dimensional renders of the swirling strength (green areas) and the SWIRL detections (blue areas) for the three magnetic field configurations at $20$ km spatial resolution. The results of SWIRL were obtained by sequentially applying the code at all the heights of the simulation. The simulations are at the same time instants as those of Fig. \ref{fig:spatial_distribution}. Vortices in SSD simulations (upper row of Fig.~\ref{fig:3d_vortices}) identified with swirling strength show elongated filaments that expand in all directions throughout the domain (similar to the results of the non-magnetic simulation of \cite{moll2012} and \cite{battaglia2021}). They can be related to shock fronts as previously noted in Sec.~\ref{sec:2d_spatial}. In the photosphere, we also observe small loops associated with vortex structures, which are mainly found around magnetic field loops. Conversely, SWIRL yields detections of small structures spread throughout the domain. They do not have large vertical extensions, and their horizontal sizes vary at different heights. They are mostly generated by hydrodynamic and magnetic baroclinic terms (Fig.~\ref{fig:ss_generation_terms}). In the photosphere, some structures are the result of vortices generated around small magnetic field loops, as also seen with swirling strength. SWIRL is able to detect some of them until they are curved with height. When the rotation leaves the horizontal plane, SWIRL no longer detects them.

Middle and lower panels of Fig. \ref{fig:3d_vortices} show the results of unipolar models. We find a different behavior compared to the SSD model. From the swirling strength renders, we observe numerous vertical filaments extending from the photosphere to the chromosphere. These structures can be associated with vertical magnetic flux tubes that expand upward. In the Bz50 model, some vortices are only found in the chromosphere, without showing connectivity with the photosphere. This indicates that vortices can be generated or sustained over time inside the chromosphere without the need for a photospheric counterpart. Moreover, some of them could be related with Alfv\'enic pulses propagating upwards \citep{battaglia2021}. This effect is absent in the Bz200 simulation. Most of the vortices detected connect the photosphere with the top of the domain, exhibiting a vertical and rigid structure. Swirling strength shows small loops in the photosphere of the Bz50 model, similar to those of the SSD simulation. They are also related to small magnetic field loops, which leads to the formation of vortices that rotate around them. In the Bz200 model, this phenomenon is barely observed. Increasing the magnetic field strength inhibits the formation of photospheric magnetic loops and consequently their associated vortex structures.

The results from SWIRL clearly show the distribution of vertically oriented vortices. We find several structures extending from the photosphere to the chromosphere, connecting different heights of the solar atmosphere. Visually, the number of vertical vortices in the Bz200 simulation is higher than in the Bz50 simulation. The stronger the magnetic field, the higher the number of vortices that are observed at the same spatial resolution. The size of these structures in both unipolar models increases with height as they are mainly located around magnetic flux concentrations. The magnetic field opens up to compensate for the pressure drop as we ascend to the chromosphere. 

Additionally, we note that vortices can bend with height. Hence, single detections of vortices in both photosphere and chromosphere should take into account this behavior, as it can give the impression of being two different structures. This is the case of the vortex located at $(x, y) \sim (5.6 \, \text{Mm}, 3.6 \, \text{Mm})$ at $Z=0$ Mm in the Bz200 model (see Fig. \ref{fig:spatial_distribution}). From the $3$D renders, we see that this structure is vertically connected with the chromospheric vortex at $(x, y) \sim (5.2 \, \text{Mm}, 3.6 \, \text{Mm})$ at $Z=1$ Mm (colored in light yellow in Fig.~\ref{fig:3d_vortices}). Despite being part of the same structure, their horizontal displacement of $300$ km between both layers makes them appear unrelated in the two-dimensional plots. 

We notice that some vortices do not originate exactly at $Z = 0$ Mm, but instead are localized slightly above the surface. Since the upper layers of unipolar simulations are characterized by low beta plasma, the dynamics of vortices with height are determined by the behavior of the magnetic field. Thus, even if the vortex is lost in the photosphere due to granulation motions, vortices can remain stable in the upper layers. In some cases, the disconnection in the lower layers can be the result of missing detections by SWIRL in the photosphere.

\subsection{Morphological properties}

We computed statistical properties of vortices obtained with both detection methods to quantify their characteristic distributions. Swirling strength only enables the measurement of the area covered by vortices, as it is not suitable for isolating individual structures. In contrast, SWIRL allows to determine individual properties of vortices, such as the vortex number density or their typical sizes as a function of height.

\begin{figure}[t]
    \centering
    % Fila 1
    \begin{subfigure}[h]{0.49\columnwidth}
        \includegraphics[width=\columnwidth]{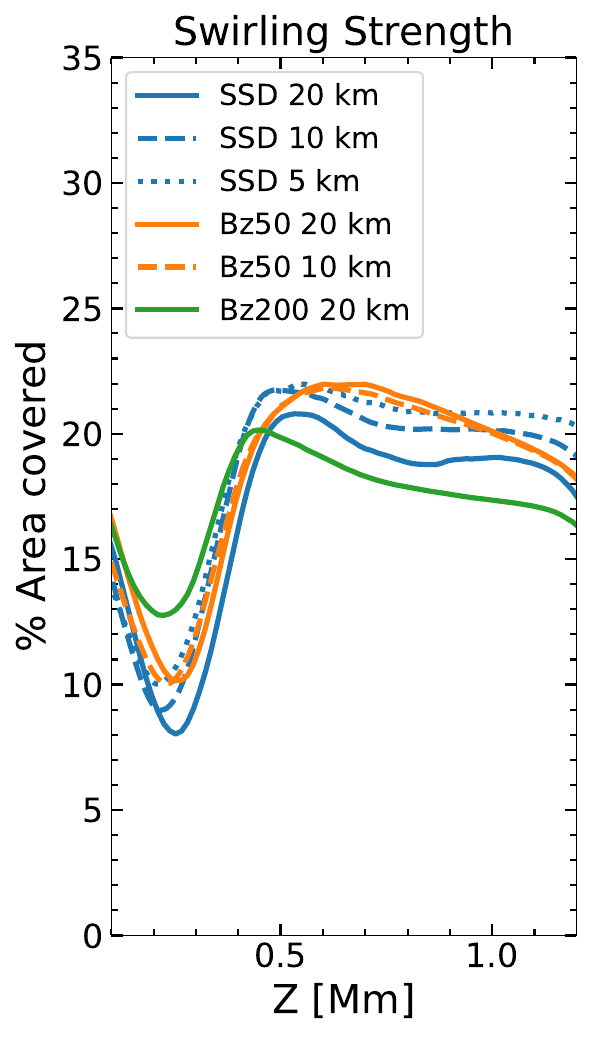}
    \end{subfigure}
    \begin{subfigure}[h]{0.49\columnwidth}
        \includegraphics[width=0.92\columnwidth]{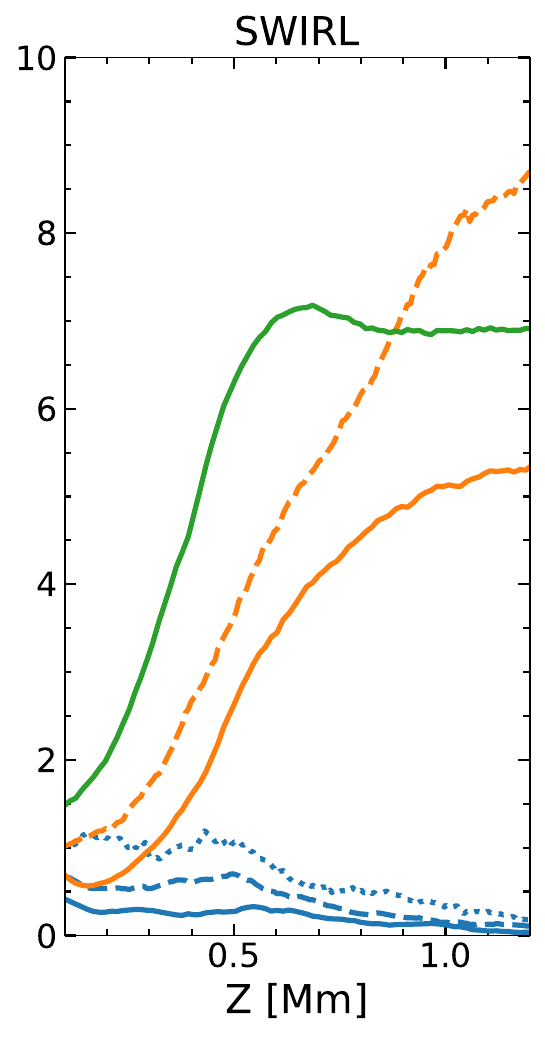}
    \end{subfigure}
    \caption{Percentage of area covered by vortices detected with swirling strength (left) and SWIRL (right) as a function of height. The meaning of the curves is indicated in the figure. Note the difference in scale of the vertical axes.}
    \label{fig:area_vortices}
\end{figure}

Figure~\ref{fig:area_vortices} shows the percentage of area covered by vortices. The left panel displays the results computed with swirling strength after applying the height-dependent threshold (given by Fig.~\ref{fig:v_rms_and_ss_var_criterion}). We observe that the area of vortices has a similar distribution in all models, regardless of the magnetic field and spatial resolution. In the photosphere, the area covered is around $10\%$, while as we ascend to the chromosphere, this quantity becomes constant close to $20\%$. Similar results have been reported in previous studies with swirling strength \citep{yadav2021, kannan2024}. 

The results from SWIRL show a very different picture (right panel of Fig.~\ref{fig:area_vortices}). They correspond to the area covered by the vortices that fulfill our filter selection (described in Sec.~\ref{sec:SWIRL}). Due to the low number of vertical vortices detected in the SSD models, the percentage of area is significantly smaller compared to the results from swirling strength. Here, the detected vortices cover less than $1\%$ of the computational domain at each height. In Bz50 models, the area covered in the photosphere is comparable to that of SSD models, and it is slightly higher in the Bz200 simulation. However, the coverage increases linearly with height. The Bz50 model reaches values of about $5\%$ for the $20$ km simulations and exceeds the $8\%$ in the $10$ km model. In the Bz200 simulation, this quantity increases up to a $7\%$. After $0.6$ Mm in height, the area covered by vortices in the Bz200 model follows a flat profile, suggesting a saturation regime. We note that spatial resolution increases the percentage of area covered by vortices.

The difference in percentage area between the detection methods highlights their relevance and illustrates how vortex analyses can be heavily impacted by the selected detection criterion. Swirling strength returns a larger number of regions associated with vortices as compared to SWIRL, as we have already noted in the two and three-dimensional plots (Fig.~\ref{fig:spatial_distribution} and Fig.~\ref{fig:3d_vortices}, respectively). SWIRL shows more localized regions that generally better fit the horizontal rotating velocity field. Besides, detections made by swirling strength also include horizontal vortices, while SWIRL is limited to vertical vortices. Thus, the percentage of area is reduced by a factor of about two in unipolar models, and more than ten times in SSD models. The large difference with SSD simulations is because most of the vortices are horizontal, while in unipolar models, horizontal structures are nearly absent since vertical magnetic fields suppress their generation.

\begin{figure*}[htbp]
    \centering
    % Fila 1
    \begin{subfigure}[b]{\columnwidth}
        \includegraphics[width=0.98\columnwidth]{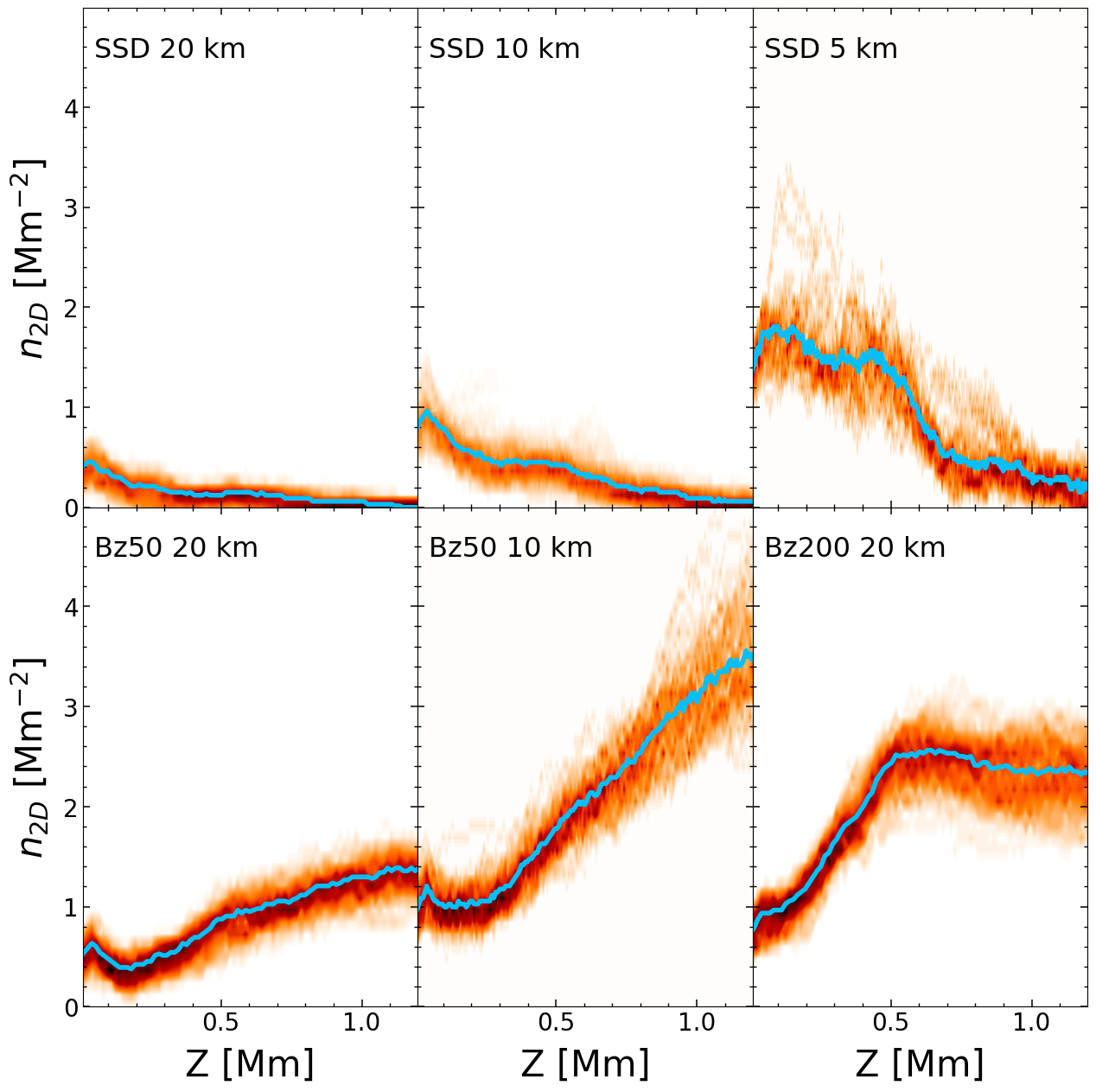}
    \end{subfigure}
    \hfill
    \begin{subfigure}[b]{\columnwidth}
        \includegraphics[width=\columnwidth]{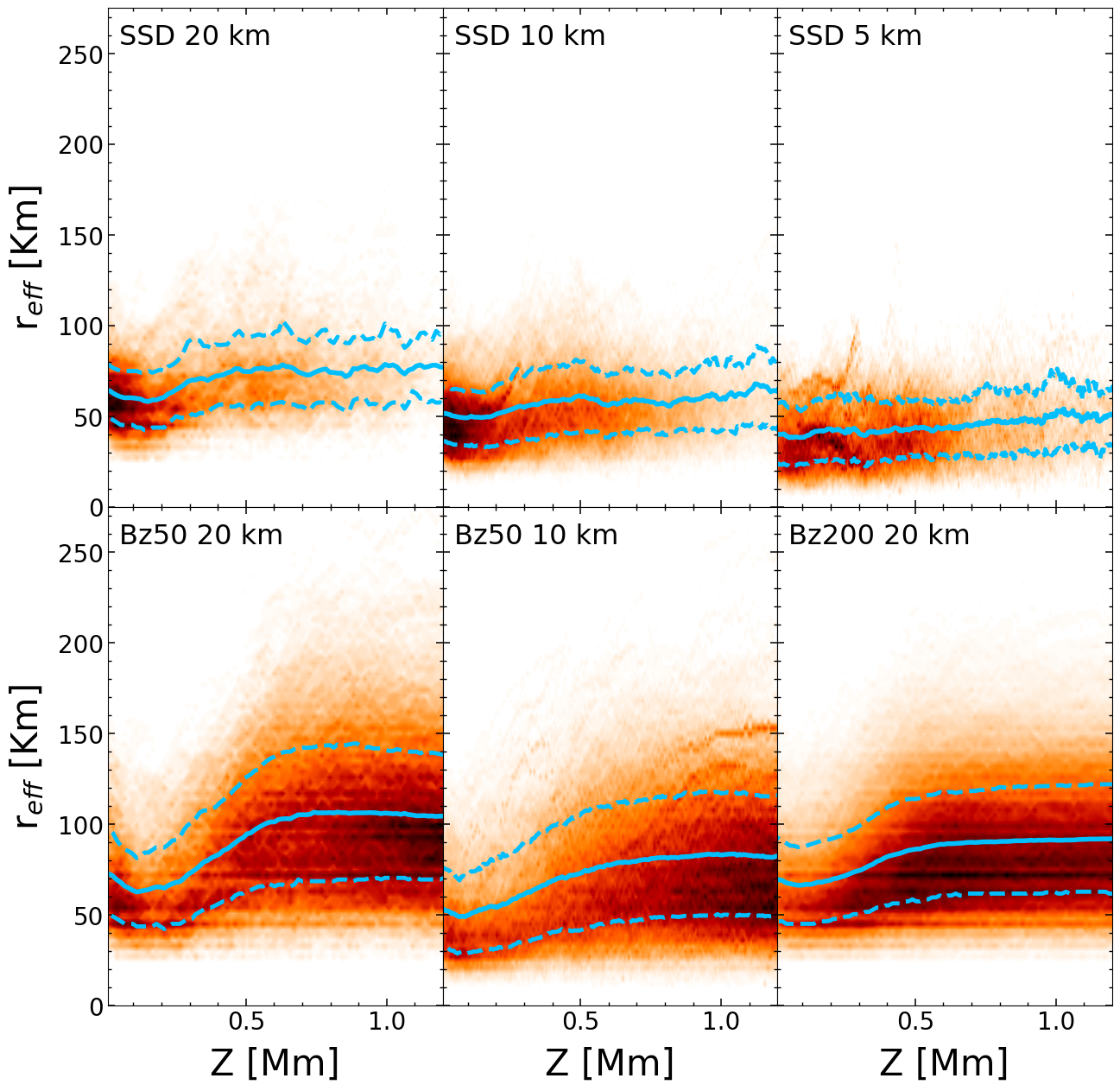}
    \end{subfigure}
    \caption{Two-dimensional histograms showing the number density (left) and estimated radius (right) of vortex detections at a given height made with SWIRL. Each data point represent the density number of vortices at a single time instant. The color scale indicates the density of data points around a given value. Darker regions indicate larger concentrations of data points. The solid blue line represents the median of the distributions and the dashed blue lines in the right panel indicate the $1\sigma$ deviation of the distributions.}
     \label{fig:density_radii_vortices}

\end{figure*}

Left panels of Fig. \ref{fig:density_radii_vortices} show two-dimensional histograms of the density number of vortices detected by SWIRL in each model as a function of height. It includes the information throughout the entire temporal evolution. The first row shows the results for the SSD models at the three available spatial resolutions. We observe a decreasing number of vortices with height. The highest number of vortices is found at the solar surface, with values around $0.4$~Mm$^{-2}$ and $1.6$~Mm$^{-2}$ in the $20$ km and $5$ km spatial resolutions, respectively. At the top of the domain, we can barely see vortical structures. The generation of vertical vortices in SSD simulations is almost absent with height due to the lack of stronger magnetic fields and their associated vertical flux tubes (as already seen in Fig.~\ref{fig:3d_vortices}). Interestingly, we note that there is an increase in the number of detections around $0.4 - 0.5$ Mm in height, which is more evident at higher spatial resolutions. At these layers, swirling strength generation terms start to increase again after the local minimum located between $0.2 - 0.3$ Mm (Fig. \ref{fig:ss_generation_terms}). In addition, the slowdown in the velocity profiles of the simulations (Fig.~\ref{fig:v_rms_and_ss_var_criterion}) occurs at the same heights, where the mechanism of energy transport changes from convection to waves. Therefore, we speculate that both effects could be related.

The second row shows the results of unipolar models. In the photosphere, the number density of vortices is higher in the Bz200 model than in the Bz50 model (as seen in Fig. \ref{fig:spatial_distribution}). We find a different behavior with height compared to the SSD results. In the Bz50 simulation, the number density grows linearly from the minimum in the solar surface to the top of the domain. In the $20$ km simulation, this value ranges from $0.4$ Mm$^{-2}$ to $1.4$ Mm$^{-2}$. This continuous increase indicates that new vortices are being generated at higher heights, which are not found at lower layers. This is related to what was observed in the three-dimensional renders of Fig. \ref{fig:3d_vortices}. Local vortices can be developed in the chromosphere without the need of a photospheric counterpart. These results are consistent with those reported by \cite{cuissa2023}, who analyzed a $50$ G model with a spatial resolution of $10$ km. This behavior changes significantly in the Bz200 simulation. Here, the number density of vortices grows linearly up to $0.6$ Mm. Thereafter, the number of vortices remains constant with number densities of about $2.5$ Mm$^{-2}$. This means that no new vortices can be generated locally in the chromosphere. The identified vortices in the upper layers originate from the same vortex structures of the photosphere. Thus, chromospheric vortices consist of long vortices connecting several layers of the solar atmosphere.

We notice that increasing the spatial resolution is the main cause of the increased number of vortices. The higher the spatial resolution, the higher the number of structures we can resolve. For a given magnetic field configuration, the dependence of the number density on atmospheric height is similar across different spatial resolutions, differing only in magnitude. Roughly, this rise is proportional to the increase in spatial resolution. For example, in the Bz50 simulation the increase in the number density of vortices between the $20$ km and $10$ km resolutions is approximately a factor of two (the SWIRL parameter file has been modified for not being affected by the change in resolution, see Appendix \ref{app:SWIRL_params}). Thus, spatial resolution is a crucial aspect when comparing vortex statistics between numerical simulations and observations or between different simulations or numerical codes.

Right panel of Fig. \ref{fig:density_radii_vortices} shows the two-dimensional histograms of the effective radius computed with SWIRL. This quantity is an estimation of the size based on the area covered by each vortex detection \citep{cuissa2023}. By assuming that structures have an almost circular shape, we computed their sizes as follows: 
\begin{equation} \label{eq:reff}
r_{\rm eff} = \sqrt{\dfrac{N_p}{\pi}} \Delta x ,
\end{equation}
where $N_p$ is the number of grid points that belong to a given vortex and $\Delta x$ is the horizontal spatial resolution of the simulation. The typical sizes observed in all SSD simulations (first row of Fig.~\ref{fig:density_radii_vortices}) vary from approximately $20$ to $100$ km in radius. Note that for the $20$ km model, the minimum radius detected is of about $30$ km. From Eq.~\ref{eq:reff}, this implies that the structure actually consists of $7$ grid points, which is above of our selection criterion (Sec.~\ref{sec:SWIRL}). The same applies for the rest of models at their respective scales. 

The distribution of sizes with height is relatively constant in SSD models, regardless of the spatial resolution. Their sizes are independent of height and are instead determined by the local hydrodynamic conditions at which they are generated. Thus, the sizes of vortices in SSD simulations follow an almost flat profile with height. 

In unipolar models, the sizes of vortices cover similar values to those of the SSD models in the photosphere. Towards the chromosphere, we note that vortex sizes increase, as they are located around vertical magnetic flux tubes which can open with height. Thus, typical sizes can reach up to $200$ km or higher in radius in the upper layers. The wider range of vortex scales in the chromosphere highlights the challenge of detecting them as the method must be able to identify velocity field curvatures at various scales. SWIRL addresses this challenge by including multiple scales (stencils) in the detection process (see Appendix \ref{app:SWIRL_params}).

We also found that vortices in the Bz50 simulation at $20$ km spatial resolution can reach larger sizes than in the Bz200 simulation at the same resolution (as also noted in Sec. \ref{sec:2d_spatial}). \citet{kannan2024} observed the same effect. It is related to the pressure balance between the inside and outside of the magnetic flux tubes where vortices are located. From the swirling strength evolution equation (Sec.~\ref{sec:ss_gen_terms}), we noted that the Bz200 model present larger contributions from the magnetic terms (baroclinic and tension) than the Bz50 model at $20$ km spatial resolution. Within the vortex structure there can be a balance between magnetic pressure and magnetic tension. The former would exert a force to expand the structure while the latter would tend to compress it. In addition, gas pressure is similar in both simulations and the magnetic pressure is larger outside vortex structures in Bz200 simulations. Therefore, the size of vortices could be more limited in stronger field simulations.

Spatial resolution also plays an important role in the horizontal extension of vortices. At higher resolutions, vortices are better resolved and new smaller structures can be detected. As a consequence, typical sizes decrease when improving spatial resolution. The decrease in vortex size is about a $20-30\%$ when the resolution is increased by a factor of two. This effect can be taken into account when comparing results with other numerical simulations or observations that utilize different spatial resolutions. By extrapolating these results, a spatial resolution of $100$ km would result in an increase of vortex sizes of about $60-70\%$ compared to the size of the vortices that we have in the $20$ km simulations. Thus, for the Bz50 simulation at $20$ km, we could obtain sizes larger than $300$ km in radius. \citet{yadav2020} degrade the spatial resolution of a simulation and also found an increase in vortex sizes.

Previous works of small-scale vortices in numerical simulations have shown a similar range of sizes. Studies using swirling strength have made estimations based on visual inspection of the detections. At photospheric layers, \citet{moll2011} reported values of around $50-100$ km in radius in dynamo simulations. \citet{yadav2021} identified structures with similar values in the photosphere and sizes of about $100-200$ km in the chromosphere in a $200$ G model. No previous works have analyzed vortices in dynamo simulations at chromospheric heights. \citet{silva2020} reported similar values using the Instantaneous Vorticity Deviation (IVD) vortex detection method in simulations up to the photosphere. Only the work by \citet{cuissa2023} has applied SWIRL to a simulation with a vertical magnetic field of $50$ G and a spatial resolution of $10$ km. For this model, we obtained comparable results across the photosphere and chromosphere. Despite the vortex size estimations are comparable between both detection methods, SWIRL allows to directly determine their extension without relying on visual estimations. 

Comparisons with observational studies are strongly affected by spatial resolution. In observations, the best spatial resolution can reach around $100$ km, while numerical simulations usually employ values between $10-20$ km. The number density of vortices reported in current observational studies in the photosphere is around $10^{-2} - 10^{-1}$ Mm$^{-2}$ \citep{vargasdominguez2011, giagkiozis2018}. In the chromosphere, number densities are about $10^{-2}$ Mm$^{-2}$ \citep{liu2019, dakanalis2022}. We observe that our results are closer in the photosphere for the $20$ km spatial resolution models, while in the chromosphere,  observational detections are between one and two orders of magnitude lower than the number of detections in any of our models. In addition, the smallest sizes detected in observational studies are of about $0.5$ Mm in diameter in the photosphere \citep{vargasdominguez2011, requerey2017, giagkiozis2018} and chromosphere \citep{park2016, liu2019}. These values represent our largest vortex sizes detected. Thus, we expect that with future higher resolution observations, vortex sizes will decrease while the number of detections increase, particularly within the chromosphere.

The magnetic field strength also influences the number and size of vortices, especially in the chromosphere (Fig.~\ref{fig:density_radii_vortices}). Most observations so far have been focused on quiet Sun regions. Within the quiet Sun, regions with relatively stronger fields - such as supergranular network - are commonly found, affecting in some extent further vortex analyses. Our results predict a larger number and size of vortices in stronger magnetized regions.

SWIRL detections focus on vertical vortices, whereas swirling strength includes both vertical and horizontal vortices oriented in any direction. Therefore, in order to make fairer comparisons, methods that detect structures oriented in the same direction should be used. Observational studies are typically limited to vertical vortices, since the velocity field is only available at a single plane perpendicular to the line of sight. Our estimations show that the area covered by vortices in the photosphere detected with swirling strength reach values above $10\%$, while SWIRL provides values below $2\%$ (Fig.~\ref{fig:area_vortices}). \citet{giagkiozis2018} estimated from observations that around $2.8\%$ of the photosphere is covered by vortices at any time. Area estimations made with SWIRL are more consistent with observations, as it only focuses on vertical vortices. However, it is important to note that the structures detected in numerical simulations are smaller and more numerous than those of observations. In the future, it might be interesting to attempt observational vortex detections at locations outside of the solar disc center, in order to provide statistics for arbitrary oriented vortices.

\subsection{Temperature distribution} 
\label{sec:temperature}

\begin{figure}[t]
    \centering
    \includegraphics[width=\columnwidth]{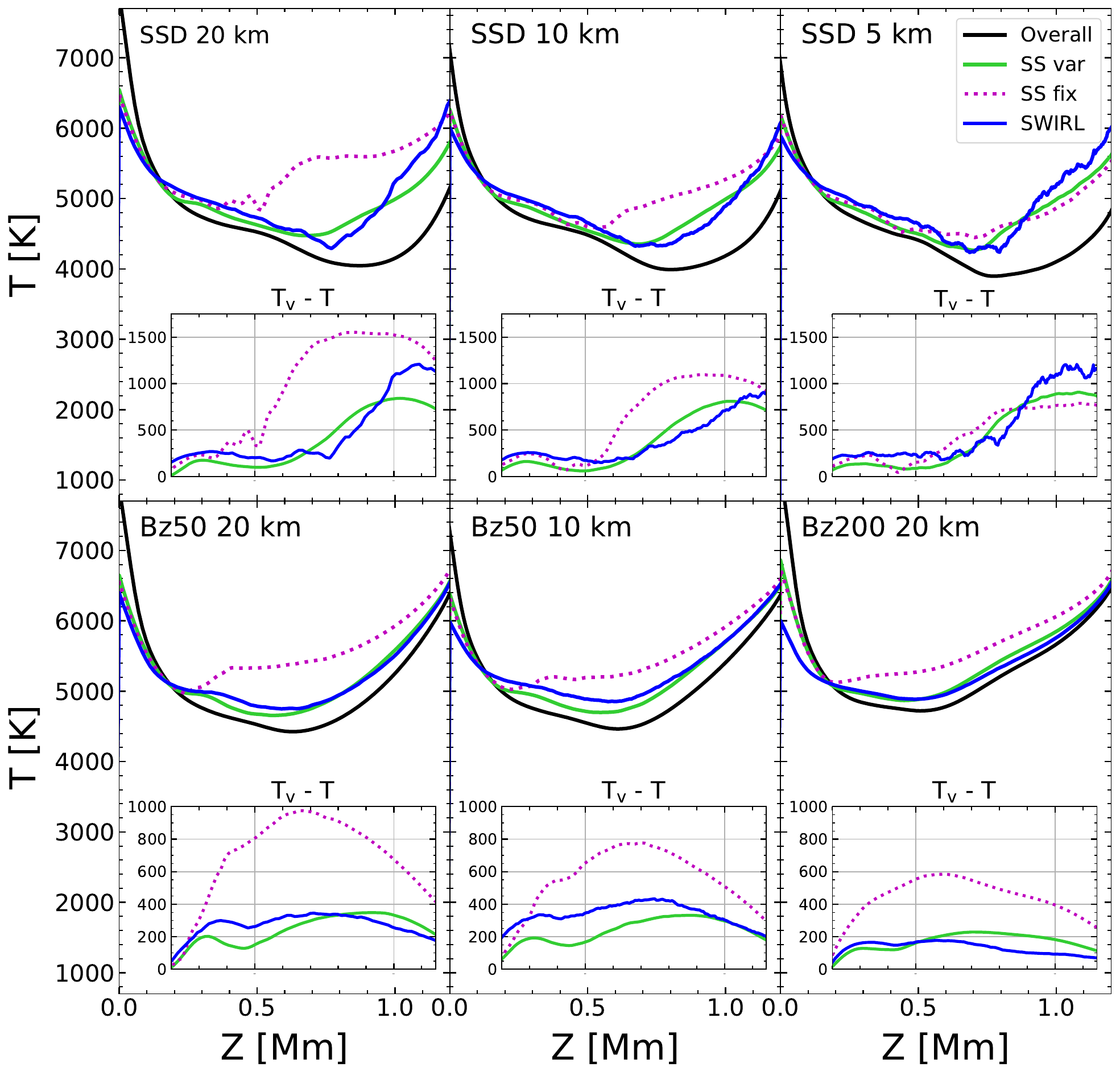}
    \caption{Horizontal and temporal average of temperature profiles in the whole computational domain (black lines) and in vortices computed with swirling strength using the height-dependent and fixed thresholds (green and purple lines, respectively) and the SWIRL code (blue lines) as a function of height. The inset plots show the temperature difference between the vortex regions and the average temperature of the simulation starting from $0.2$ Mm. Note the different vertical scale between both rows.} 
     \label{fig:temperature}
\end{figure}

Figure \ref{fig:temperature} shows the spatio-temporal averages of the temperature profiles in the whole computational domain (black lines) and in vortex regions computed with swirling strength (green and purple lines) and SWIRL (blue lines) as a function of height. Two independent measurements are illustrated for the swirling strength. Green lines correspond to the averages computed using the height-dependent criterion, while purple lines contain the averages using a fixed threshold for the entire computational box. For the fixed threshold, we selected locations with swirling strength periods lower than $100$ s ($\lambda_{\mathrm{ci}} > 0.0628$ Hz) for unipolar models. We chose this value solely to compare our results against previous studies that analyze unipolar models \citep[e.g.,][]{yadav2021}. However, according to Fig. \ref{fig:v_rms_and_ss_var_criterion}, this threshold is very restrictive compared to typical average values for swirling strength, leading to the isolation of the most intense vortex cores. For SSD models we selected regions with swirling strength periods lower than $200$ s ($\lambda_{\mathrm{ci}} > 0.0314$ Hz). This value lies between the photospheric values for the three available resolutions and serves as an estimate for SSD models. We have not found previous studies that apply a fixed threshold to SSD simulations.

Temperatures computed with swirling strength using the height-dependent threshold and the SWIRL code show qualitatively similar behavior, but with differences in magnitude. We observe that, in the photosphere, the temperature of vortices is below the average temperature profile of the simulation in all the analyzed models. This is because vortices are mainly located inside intergranular lanes (as seen in the first row of Fig. \ref{fig:spatial_distribution}), which are cooler than their surroundings. This behavior extends up to a height of $0.2$ Mm, where the temperature distribution is reversed and vortices become hotter than their surroundings. Then, vortices remain hotter than the average temperature profile of the simulations up to the top of the domain.

We find that the vortices from all the spatial resolutions exhibit similar temperature profiles. Although the number and size of the vortices change as the spatial resolution increases, temperature profiles are unaffected. These findings support the idea of \citet{yadav2020}, who indicated that large vortices at low resolution actually consist of several small vortices within them. Consequently, the temperature distribution between spatial resolutions is similar.  

Temperature profiles of vortices are also similar between different magnetic field configurations. We note that the main differences are actually related with the increase of the average temperature profile of each model (black lines). For example, the average temperature of the Bz200 simulation is significantly higher than the temperature of the Bz50 and SSD models \citep[as reported by][]{khomenko2025}. Thus, vortices could share the same mechanism responsible for the temperature increase, independently of the magnetization of the model.

The inset plots of Fig. \ref{fig:temperature} show in detail the temperature difference between vortices (computed with both detection methods) and the average temperature profiles starting from $0.2$ Mm. From the results of the height-dependent criterion and SWIRL, we observe that the temperature increase of vortices in SSD model is about $100 - 200$ K up to $0.6$ Mm in height. Then, the temperature of vortices significantly rises, reaching differences higher than $1000$ K compared to the mean temperature profiles. 

Unipolar models show different temperature profiles. The temperature increase in vortices remains roughly constant compared to the average temperature in the atmosphere from about $0.2$ Mm to $1$ Mm above the surface. This increased temperature is higher in the Bz50 simulations, reaching differences of about $200-400$ K, while for Bz200 model is only $200$ K above the average profile. Therefore, all models show temperature enhancements of at least $200$ K by using the height-dependent criterion and the SWIRL code up to $1$ Mm. 

The results with a fixed threshold in swirling strength (purple dotted lines) exhibit larger temperature differences than the height-dependent criterion and SWIRL. In SSD and Bz50 models, the difference in temperature reduces as the spatial resolution is improved. This is a consequence of the changes in the typical values of swirling strength with spatial resolution (Sec. \ref{sec:ss_threshold}). For the Bz200 simulation, we observe temperature difference reaching $400-500$ K, which is twice the values obtained with the height-dependent criterion and SWIRL. \cite{yadav2021} computed the temperature profile of vortices in a $200$ G simulation at $10$ km spatial resolution. They applied swirling strength using the same fixed threshold that we employed for Bz200 model. They found a temperature increase of about $300-400$ K in vortices, which remains approximately constant up to $2.3$ Mm above the solar surface. Despite using different numerical codes, we obtained similar results when applying the same method. The small difference in temperature can be related to the different spatial resolution of the models. The stratification of the vortex temperature depends on the choice of detection method and how it is applied. This highlights the importance of using methods that do not rely on visual inspection, as the height-dependent criterion of swirling strength or the SWIRL code. 

In the upper part of the domain of unipolar models, we note that the temperature of the vortices converge towards the average temperature of the simulation. This may be the result of effects produced by the top boundary condition or by processes involving plasma ionization inside vortices. To verify this, a more detailed analysis of the effect of boundary conditions and the energy transport and dissipation has to be done. Note, however, that SSD simulations do not show this effect in the top boundary layer.

\section{Conclusions} \label{sec:conclusions}

In this paper, we present the first analysis of small-scale vortices in a set of realistic MHD numerical simulations covering different magnetizations and spatial resolutions. We compared two vortex detection methods: the swirling strength criterion \citep{zhou1999} and the SWIRL code \citep{cuissa2022}. 

We found SWIRL to be a robust tool for the detection of vertical vortices, enabling both comprehensive statistical studies and single-vortex analysis. In works employing swirling strength, we suggest to apply a height-dependent threshold since it is highly dependent on the local velocity field. 

Our analysis confirms that magnetic fields significantly affect vortical structures, promoting the formation of vertical vortices, particularly at chromospheric heights. The generation of vortices is mainly driven by magnetic contributions in stronger field simulations, whereas hydrodynamic forces dominate the formation of vortices in weaker fields. Interestingly, magnetic pressure could play an important role in weaker fields models. Intermediate magnetic fields suggest that new vortices can be formed at higher layers without a photospheric counterpart, supporting the idea of Alfv\'enic pulses propagating upwards \citep{battaglia2021}. Vortices in stronger fields require a photospheric connection, indicating that these environments could favor the development of extended structures that act as conduits for energy and wave propagation. Vortex sizes increase with height due to the expansion of magnetic field lines in unipolar models. Consequently, we expect in observations a higher number of vortices in regions with intermediate and strong magnetic fields, increasing their sizes with height.

Higher spatial resolution allows to detect a higher number of vortices with smaller sizes. However, we suggest that the generation of smaller vortices may be limited to the characteristic scales of physical viscosity. Therefore, spatial resolution may introduce differences when comparing results with other numerical simulation or observational studies. Moreover, to ensure a direct comparison with observational studies, vortex detection methods should be focused exclusively on vertical structures, as observations are typically limited to the analysis of horizontal velocity fields in a single plane. We expect that new high-spatial resolution observations show a close agreement with the morphological vortex distributions reported in this study. 

Temperature profiles show that vortices have higher temperatures than the average profiles of the atmosphere in all of our models from approximately $0.2$ Mm. The temperature enhancement depends on the magnetic field configuration, while we do not observe differences among spatial resolutions. All models show an increase of at least $200$ K. Nevertheless, a detailed analysis of the energy dissipation mechanism is necessary to fully understand these temperature differences.

\begin{acknowledgements} 
The authors acknowledge the support from Agencia Estatal de Investigación del Ministerio de Ciencia e Innovación and the European Regional Development Fund (ERDF "A way of making Europe") through grants PID2021-127487NB-I00 and PID2024-156538NB-I00; and from the OSCARS project No. 01-454 "Federation of Solar Data (FSD)". MKP acknowledges the grant PRE2022-103940, funded by MCIN/AEI/10.13039/501100011033 and by ESF+. TF acknowledges grant CNS2023-145233 funded by MICIU/AEI/10.13039/501100011033 and by “European Union NextGeneration EU/PRTR” and grant RYC2020-030307-I funded by MCIN/AEI/ 10.13039/501100011033 and by “ESF Investing in your future”. The authors thankfully acknowledge the technical expertise and assistance provided by the Spanish Supercomputing Network (Red Española de Supercomputación), as well as the computer resources used: LaPalma Supercomputer, located at the Instituto de Astrofísica de Canarias, and MareNostrum based in Barcelona/Spain.
\end{acknowledgements}

\bibliographystyle{aa} 
\bibliography{vortices}

\begin{appendix}

\section{SWIRL parameters}
\label{app:SWIRL_params}

Detecting vortices using SWIRL requires modifying a set of parameters to obtain accurate results. Table~\ref{tab:specific_params} contains the main parameters we have adjusted (stencil and noise) in all the models. We determined these parameters by comparing the results with the horizontal velocity field. We first adjusted the parameters in the lower resolution simulations, and then we extended the results to the higher resolution models. The primary parameter to be modified is the stencil, as it modifies the scale at which the velocity field curvature is detected. This means that a stencil of $1$ performs the calculations in all grid cells, while a stencil of $5$ is applied every $5$ grid points. In this way, higher stencil values allow for the detection of larger-scale rotations, whereas smaller stencils focus on the detection of small-scale motions. In simulations where we increase the resolution by a factor of two, we increased the stencils used in the lower-resolution simulations by the same factor to keep a comparable size of the detected structures, avoiding potential effects introduced by resolution in the subsequent results. Table~\ref{tab:general_params} contains the rest of parameters that are common for all the simulations analyzed. Table~\ref{table:swirl_params} summarizes the meaning of all the parameters available, although further details on the parameters can be found in \citet{cuissa2022}.

\renewcommand{\arraystretch}{1.4} 
\begin{table}[h!]
\caption{List of specific parameters of SWIRL applied to each simulation.}
\centering
\begin{tabular*}{0.9\columnwidth}{@{\extracolsep{\fill}}lcc}
\hline\hline
Simulation & Stencil & Noise parameter \\
\hline
SSD $20$ km   & $[1,2,3]$           &  $1.2$  \\
SSD $10$ km   & $[1,2,3,4,6]$       &  $1.2$  \\
SSD $5$ km    & $[1,2,3,4,6,8,12]$  &  $1.2$  \\
Bz50 $20$ km  & $[2,3,4,6,7]$       &  $1.1$  \\
Bz50 $10$ km  & $[2,3,4,6,8,12,14]$ &  $1.1$  \\
Bz200 $20$ km & $[2,3,4,6,7,10]$    &  $1.1$  \\
\hline
\end{tabular*}
\label{tab:specific_params}
\end{table}
\renewcommand{\arraystretch}{1}

\renewcommand{\arraystretch}{1.4} 
\begin{table}[h!]
\caption{List of common parameters of SWIRL applied to all simulations.}
\centering
\begin{tabular*}{0.6\columnwidth}{@{\extracolsep{\fill}}ll}
\hline\hline
[$\epsilon_{\lambda}, \kappa_{\lambda}, \delta_{\lambda}$] & 
[$0.0, 0.9, 0.9$] \\
d$_{c}$   & $3$           \\
Adaptive d$_{c}$   & False       \\
Fast clustering   & True       \\
Kernel   & Gaussian     \\
Decision method  & $\gamma$    \\ 

[p$_{\rho}$, p$_{\delta}$, p$_{\gamma}$] & [$0.9$, $0.9$, $1.01$] \\
Kink parameter & $0.5$ \\
\hline
\end{tabular*}
\label{tab:general_params}
\end{table}
\renewcommand{\arraystretch}{1}

\begin{table}[h!]
\caption{Brief description of SWIRL parameters.}
\label{table:swirl_params}
\centering
\renewcommand{\arraystretch}{1.3}
\begin{tabular}{lp{0.66\columnwidth}}
\hline\hline

Parameter & Description \\
\hline 
\textit{Stencils} & Defines the grid spacing used to compute spatial derivatives. Multiple stencils help to detect large and small scale vortices. \\

\textit{Noise parameter} &  Removes grid points located further from the vortex center than this factor times the estimated effective radius ($r_{eff}$). \\

\textit{($\epsilon_{\lambda}, \kappa_{\lambda}, \delta_{\lambda}$)} & Thresholds used in the code to filter out negligible swirling strength or Rortex values ($\epsilon_{\lambda}$), and to discard regions where the radial velocity of the flow dominates over the rotational velocity ($\kappa_{\lambda}, \delta_{\lambda}$). \\

\textit{$d_c$} & Parameter used to compute the critical distance used in the clustering algorithm to evaluate the local density of points. \\

\textit{Adaptive $d_c$} & Automatically calculates the optimal $d_c$. \\

\textit{Fast clustering} & Activates the grid-adapted version of the fast clustering CFSFDP algorithm \citep{rodriguez2014}, significantly reducing the computational costs for large datasets. \\

\textit{Kernel} & Mathematical function used to compute local densities in the clustering algorithm. It can use a strict cutoff boundary or a Gaussian function that decays exponentially. \\

\textit{Decision method} & Select the specific strategy applied to select true vortex centers ($\rho-\delta$ or $\gamma$). \\

\textit{(p$_{\rho}$, p$_{\delta}$, p$_{\gamma}$)} & List of parameters used in the clustering  algorithm to select the cluster centers. \\

\textit{Kink parameter} & Filters out false vortices caused by partial rotations of the flow.\\

\hline
\end{tabular}
\end{table}
\renewcommand{\arraystretch}{1}
%%%%%%%%%%%%%%%%%%%%%%%%%%%%%%%%%%%%%%%%%%%%%%%%%%%%%%%%%

\section{SWIRL Rortex filter}
\label{app:SWIRL_Rortex_filter}

In Section \ref{sec:SWIRL}, we filtered the SWIRL results using the Rortex quantity. The aim of this process is to remove vortices with very low rotational speeds, since their contribution in terms of energy is small compared to faster vortices. This process is similar to what is done when selecting a threshold in the swirling strength method to discard weak vortices. We used the Rortex value computed with the smallest stencil in each model to identify strong rotations in both large and small vortices, as it focuses on smaller scales. Using larger stencils would favor only the detection of larger structures. In order to avoid excluding large vortices that may have higher Rortex contributions in other stencils, we require that at least $10\%$ of the area of each vortex has Rortex values above the mean value at each height when using the smallest stencil.

\begin{figure}[t]
    \centering
    \includegraphics[width=\columnwidth]{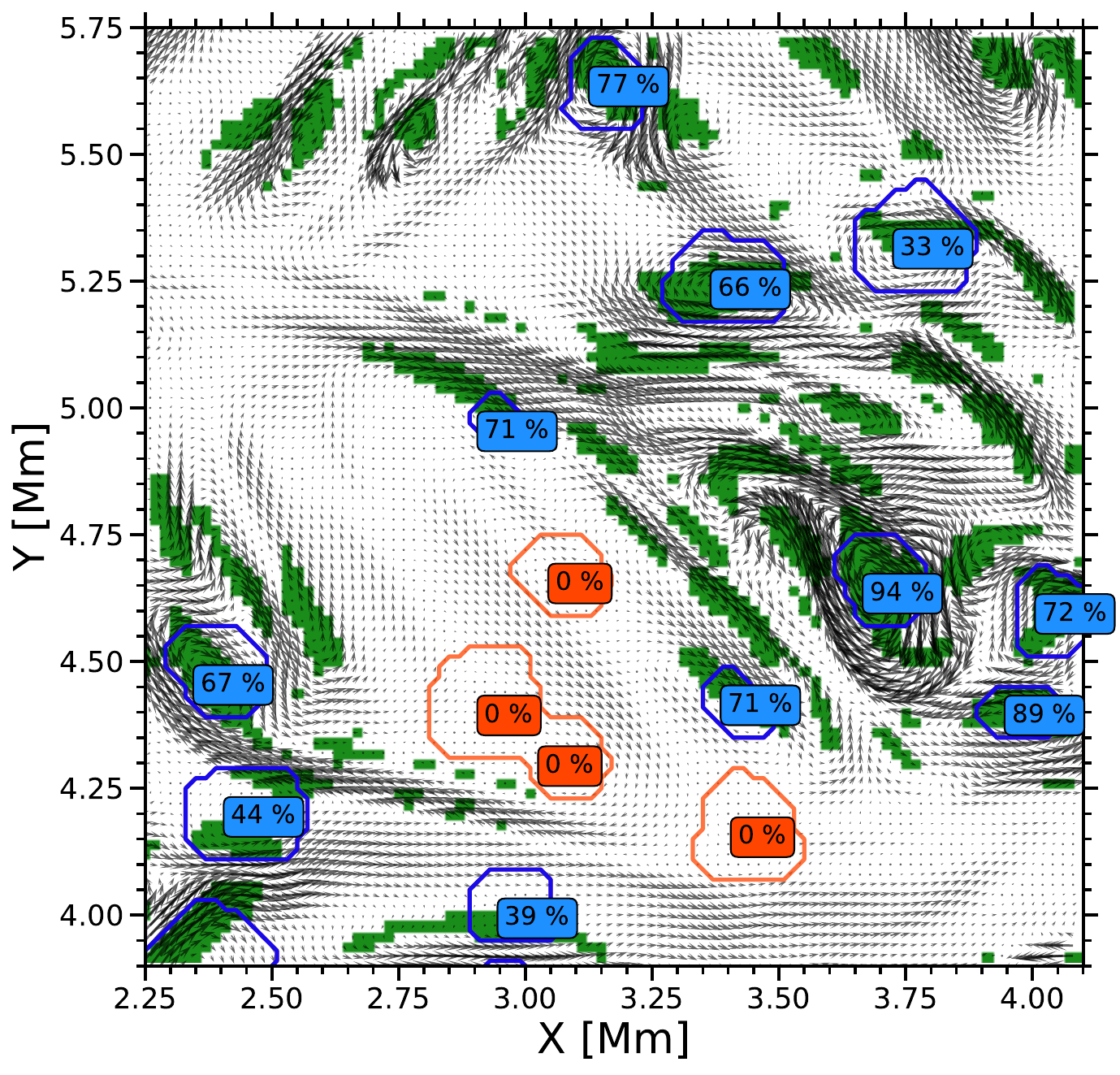}
    \caption{Vortex detections obtained with SWIRL in a zoomed-in region of the chromosphere ($Z=1$ Mm) for the Bz200 model at a $20$ km resolution. The snapshot corresponds to the same time instant as in Fig. \ref{fig:spatial_distribution}. Orange contours indicate detections removed by the minimum Rortex area criterion, while blue contours show the validated detections. The percentage values within each vortex represent the fraction of grid points with Rortex values exceeding the average magnitude at this height. Black arrows represent the horizontal velocity field ($v_x$, $v_y$).} 
     \label{fig:Rortex_vortices_z1}
\end{figure}

Fig. \ref{fig:Rortex_vortices_z1} shows a close-up view of a slice at $Z=1$ Mm for the Bz200 model at the same time instant shown in Fig. \ref{fig:spatial_distribution}. Here, the orange contours show examples of detections that have been filtered out due to the Rortex criterion, while blue contours are vortices that satisfy this criterion. The numbers inside each contour indicate the percentage of area covered by the Rortex quantity exceeding the average value at this height. Thus, we can see that the orange contours contain no values above the average ($0 \%$). Comparing these structures with the velocity field, we see that their velocity field is slow compared to the other structures, and it is not easy to determine whether the velocity field is actually rotating. All the blue contours show a stronger rotating velocity field, with higher area percentage values.

\section{Horizontal vortices}
\label{app:horizontal_vortices}

\begin{figure*}[h!]
    \centering
    \includegraphics[width=\textwidth]{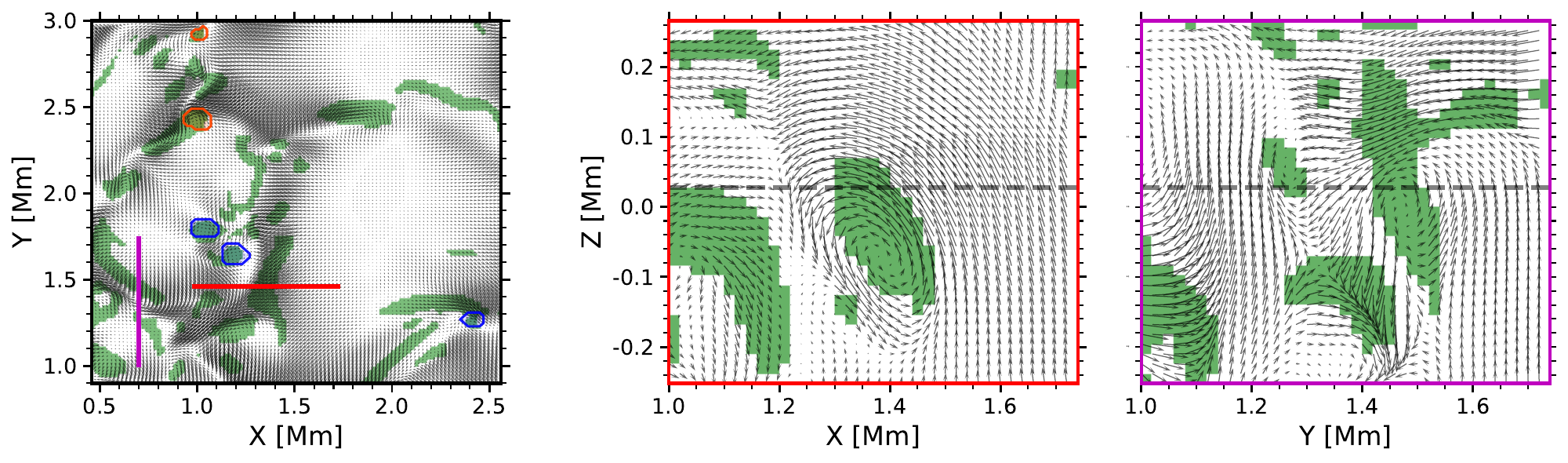}
    \caption{Horizontal vortices in the SSD model. The left panel shows the same close-up view as in Fig. \ref{fig:spatial_distribution} for the SSD model at $Z=0$ Mm, with two overplotted lines (red and magenta) crossing different regions of swirling strength values (green patches). The middle and right panels show vertical slices taken along the positions of the lines in the left panel. Black arrows show the velocity field projected onto each slice, i.e., $(v_x, v_z)$ for the middle panel (red line) and $(v_y, v_z)$ for the right panel (magenta line). The horizontal black dashed line indicates the reference height ($Z=0$ Mm) of the horizontal slice shown in the left panel.} 
     \label{fig:Horizontal_vortices_z0}
\end{figure*}

\begin{figure*}[h!]
    \centering
    \includegraphics[width=\textwidth]{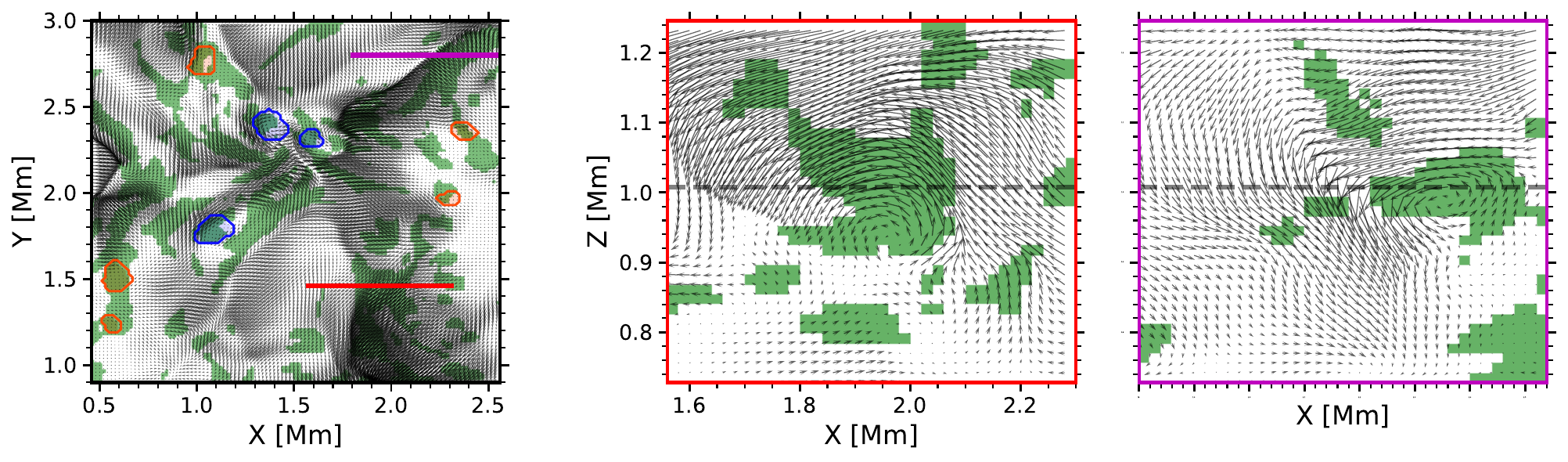}
    \caption{Same as in Fig. \ref{fig:Horizontal_vortices_z0} but at $Z = 1$ Mm.} 
     \label{fig:Horizontal_vortices_z1}
\end{figure*}

In Fig.~\ref{fig:spatial_distribution} we noted how the swirling strength method was able to detect elongated regions where the horizontal velocity field does not follow a rotational pattern. These regions were observed in the intergranular lanes of all models, regardless the magnetization, as well as in the chromosphere of the SSD simulations. Figs.~\ref{fig:Horizontal_vortices_z0} and \ref{fig:Horizontal_vortices_z1} show vertical slices of these regions for the SSD simulation in the photosphere and chromosphere, respectively, at the same location shown in the zoom-in regions of Fig.~\ref{fig:spatial_distribution}. Here, we see that these regions clearly exhibit a rotating velocity field along the horizontal axis, confirming their horizontal nature.

\end{appendix}

\end{document}